\documentclass[prd,preprint,tightenlines]{revtex4}

\usepackage{amsmath,amssymb}
\usepackage{bm} 

\usepackage{color} 
\usepackage{graphicx}

\usepackage[colorlinks]{hyperref} 
\hypersetup{
	citecolor=blue,
	linkcolor=blue,
	urlcolor=blue
}

\usepackage{epstopdf}
\usepackage{comment}

\usepackage{multirow}

\newcommand{\fig}[1]{Fig.~\ref{#1}}

\newcommand{\tab}[1]{Tab.~\ref{#1}}

\newcommand{\eq}[1]{Eq.~(\ref{#1})}

\usepackage{mfirstuc} 
\renewcommand{\paragraph}[1]{} 

\newcommand{\beq}[1]{\begin{equation}\label{#1}}
\newcommand{\eeq}{\end{equation}}
 \newcommand{\bea}[1]{\begin{eqnarray}\label{#1}}
 \newcommand{\eea}{\end{eqnarray}}
 \newcommand\figcaption{\def\@captype{figure}\caption}
 \newcommand\tabcaption{\def\@captype{table}\caption}
\newcommand{\PYTHIA}{{\tt PYTHIA~8.2}}
\newcommand{\EPOSLHC}{{\tt EPOS-LHC}}
\newcommand{\DPMJET}{{\tt DPMJET-}III}
\newcommand{\GALPROP}{{\tt GALROP-v54}}

\newcommand{\pHebar}{p_0^{\scriptscriptstyle\overline{\mathrm{He}}}}

\newcommand{\pDbar}{p_0^{\scriptscriptstyle\overline{\mathrm{D}}}}
\newcommand{\Hebar}{^3\overline{\text{He}}}

\begin{document}
\title{Prospects of detecting  dark matter through cosmic-ray antihelium \\
 with the  antiproton constraints }
\author{
 Yu-Chen Ding$^{1,2}$\footnote{dingyucheng@itp.ac.cn}
 Nan Li$^{1,2}$\footnote{linan2016@itp.ac.cn}, 
 Chun-Cheng Wei$^{1,2}$\footnote{ccwei@itp.ac.cn},
 Yue-Liang Wu$^{2,1}$\footnote{ylwu@itp.ac.cn} and 
 Yu-Feng Zhou$^{1,2}$\footnote{yfzhou@itp.ac.cn}}
 \affiliation{$^{1}$CAS Key Laboratory of Theoretical Physics, 
Institute of Theoretical Physics, Chinese Academy of Sciences, Beijing 100190, 
China.\\
$^{2}$University of Chinese Academy of Sciences, Beijing 100049, China.}

 \begin{abstract}
Cosmic-ray (CR) antihelium is  an important observable for dark matter (DM) 
indirect searches  due to  extremely low secondary backgrounds towards low 
energies.
In a given DM model, the predicted  CR antihelium flux is expected to be  
strongly  correlated  with that of CR antiprotons.  In this work, we use the AMS-02 
$\bar p/p$ data to constrain the DM annihilation cross section, and the ALICE data on 
the $\Hebar$ and $\overline{\text{T}}$ productions   to determine 
the  parameters in the coalescence model  for anti-nucleus formation.
The hadronic cross sections are estimated using Monte-Carlo event generators
including {\tt EPOS-LHC} and {\tt DPMJET}.
Based on these constraints, we make predictions for the maximal antihelium flux  
for typical  DM annihilation final states, and perform a detailed analysis on  the 
uncertainties due to  the DM density profiles and CR propagation models.  
We find that the  results are  highly insensitive to both of them,
but still significantly depend on the  hadronization models in event generators.
The prospects of detecting antihelium for the AMS-02 experiment is discussed.
We show that with very optimistic assumptions, CR $\Hebar$ is within the reach  of 
the AMS-02 experiment. The $\Hebar$ events which can be detected by AMS-02  are likely to have kinetic energy  $T \gtrsim 30$~GeV, which is consistent with 
the preliminary AMS-02 search results. The events which can be observed by AMS-02
are likely to arise dominantly from secondary backgrounds rather than DM interactions.
 \end{abstract}
\date{\today}
 \maketitle

\section{Introduction}
Although the existence of dark matter (DM) as the dominant component  of matter in the present-day Universe has been well established by observations, the particle nature of DM remains largely unknown.  If DM particles in the Galactic halo can annihilate or decay into the standard  model (SM) stable final states, they can make extra contributions to the fluxes of cosmic-ray (CR) particles, which can be  probed by high precision DM indirect search experiments.  
Among many CR observables, CR antimatter, such as CR positrons, antiprotons 
and heavier anti-nuclei such as antideuteron and antihelium are considered to be 
relatively rare as they are dominated by CR secondaries produced by the collisions 
of primary CR particles onto the interstellar gas.  Thus CR antiparticles are 
expected to be sensitive to extra contributions and can be important probes of DM 
interactions.

In recent years, a number of experiments including AMS-02 have confirmed an 
unexpected rise in the CR positron flux above $\sim$10~GeV%
~\cite{Beatty:2004cy,Adriani:2008zr,FermiLAT:2011ab,Accardo:2014lma}.
DM annihilation or decay can be a possible explanation to this phenomena 
(see e.g. Refs.%
~\cite{Kopp:2013eka,%
Bergstrom:2013jra,%
Ibarra:2013zia,%
Jin:2013nta}
for discussions related to the AMS-02 data) which is, however, subject to stringent constraints such as that from the observations of $\gamma$-rays from dwarf galaxies~\cite{Fermi-LAT:2016uux}, the Galactic center~\cite{Abdallah:2016ygi}, 
and the measurement of anisotropy in the cosmological microwave background 
(CMB)~\cite{Ade:2015xua}.
Another important observable is the CR antiproton which has been
measured by a number of experiments such as 
PAMELA~\cite{Adriani:2012paa}, BESS-polar II~\cite{Abe:2011nx} and 
AMS-02~\cite{Aguilar:2016kjl}.  The high precision measurement of AMS-02 
shows that in a large rigidity range from $\sim 1$ to 450~GV, the antiproton flux is 
in an overall agreement with the secondary origin of CR antiprotons, which can be 
used to place stringent constraints on the properties of DM particles
(see e.g. Refs.%
~\cite{
1504.04276,%
Jin:2015sqa,%
Lin:2016ezz,%
Reinert:2017aga}).

Despite  tiny production rates, heavier anti-nuclei such as antideuteron 
($\overline{\textrm{D}}$) and antihelium-3 ($^3\overline{\textrm{He}}$) can also be 
important probes of DM, and can be searched by the experiments such as
BESS~\cite{Abe:2012tz}, 
AMS-02~\cite{Giovacchini:2007dwa,Kounine:2010js} 
and GAPS~\cite{Aramaki:2015laa}.
With the increase of the atomic mass number $A$, the fluxes of anti-nuclei
are expected to decrease rapidly due to smaller volume of phase space
for the formation of anti-nuclei for both DM annihilation and
secondary production.  
However, the fluxes of secondary anti-nuclei are further suppressed in the 
low-energy region as the final state anti-nuclei are highly boosted due to  the  
high production  thresholds in $pp$-collisions (17$m_{p}$ for 
$\overline{\textrm{D}}$ and 31$m_{p}$ for $^{3}\overline{\text{He}}$, where 
$m_{p}$ is the proton mass). The extremely low background makes it easier to 
single out the DM contributions in the low energy region below $\sim$10~GeV. 
Furthermore, at very high energies the secondary production is suppressed by  the 
rapid falling of the primary CR  flux at high energies (the CR proton flux scales 
with energy $E$ as $E^{-2.75}$).

\paragraph{This work}
CR antideuteron production from DM interactions has been extensively discussed
(for a recent review, see e.g.~\cite{Aramaki:2015pii}).  
The CR antihelium production was first discussed in%
~\cite{Carlson:2014ssa,Cirelli:2014qia}.
It has been noticed that in a given DM model, the predicted fluxes of antiproton
and antideuteron should be correlated. Thus the constraints from the
antiproton data can be used to set limits on the maximal fluxes of
antideuteron%
~\cite{Ibarra:2012cc,Fornengo:2013osa,Lin:2018avl}.
The same strategy can be applied to the case of CR antihelium
production, which was briefly discussed in~\cite{Herms:2016vop} based
on a fixed DM profile and CR propagation model and the value of
coalescence momentum $p_0$ in the coalescence model for antinucleon
formation inferred from rescaling the value for antideuteron.

In this work, we perform an updated analysis on  the prospects of detecting 
$^{3}\overline{\text{He}}$ events in the AMS-02 experiment, motivated partly by 
the recent progresses  in searching for heavier anti-nuclei made by 
AMS-02~\cite{Ting:2018}.
We use the AMS-02 $\bar p/p$ data to constrain the DM annihilation cross 
sections, taking into account the uncertainties in  DM profiles and CR 
propagation models, and use the ALICE  antinuclei production data from 
$pp$-collisions to directly constrain  $p_0$ for various Monte-Carlo (MC) 
hadronization event generators including {\tt PYTHIA}, {\tt EPOS-LHC}
and {\tt DPMJET}. Based on  these constraints, we make predictions for the 
maximal antihelium flux  for typical  DM annihilation final states.
We find that the resulting predictions for the maximal fluxes of  
$^{3}\overline{\text{He}}$ are highly insensitive to the choice of DM profiles 
and propagation models, due to the fact that the variation in the DM 
density profile and propagation model mainly leads to a rescaling of the DM 
annihilation cross sections in such a way that the same antiproton flux is 
reproduced. The results, however, are still significantly depends on the 
hadronization models in the  MC event generators.
We show that with very optimistic estimations of detection efficiency and 
acceptance, CR $\Hebar$ is within the sensitivity of the AMS-02 experiment with 
a whole lifetime of data taking. Furthermore, We find that  the events which can be 
detected by AMS-02 first are likely to have kinetic energy  $T \gtrsim 30$~GeV, 
which is consistent with the very preliminary AMS-02 antihelium 
measurements~\cite{Ting:2018}. However, they should dominantly arise from
secondary backgrounds rather than DM annihilation.

This paper is organized as follows: 
In section~\ref{sec:model}, we give a brief overview of the coalescence model for 
the formation of heavy anti-neuclei.
In section~\ref{sec:ALICE}, we use the ALICE  antinuclei production data from 
$pp$-collisions to directly constrain  $p_0$ for various MC event generators 
\PYTHIA, \EPOSLHC~and \DPMJET~for hadronization, and discuss the energy 
spectra of $\Hebar$ in the cases of DM annihilation and $pp$-collisions.
In section~\ref{sec:propagation}, we constrain the DM annihilation
cross section using the AMS-02 $\overline{p}/p$ ratio data. 
In section~\ref	{sec:prospects}, we 
calculate the $^3\overline{\textrm{He}}$ flux at the top of atmosphere and discuss 
the detection prospect of the AMS-02 experiment. 
The conclusions of this work are summarized in section~\ref{sec:conclusion}.

\section{The coalescence model}\label{sec:model}
\paragraph{The model}
We adopt the coalescence model%
~\cite{Butler:1963pp,Schwarzschild:1963zz,Csernai:1986qf}
to describe the formation of  an anti-nuclei $\bar{\mathrm{A}}$ from anti-nucleons.  
In this model a single parameter of coalescence momentum 
$p^{\bar{\mathrm{A}}}_0$ is introduced to determine whether the anti-nucleons 
produced in a collision process can merge into an
anti-nucleus. The basic assumption is that the anti-nucleons are able to
merge into an anti-nucleus only if a proper combination of the
relative four-momenta of the constituent nucleons is less than 
$p^{\bar{\mathrm{A}}}_0$.  For instance, in the case of antideuteron formation, the 
coalescence criterion
is defined as 
\beq{Dbar1}
||k_{\bar{p}}-k_{\bar{n}}|| = \sqrt{(\Delta \vec{k})^2-(\Delta E)^2} <\pDbar, 
\eeq%
where $k_{\bar{p}}$ and $k_{\bar{n}}$ are the four-momenta of $\bar{p}$ and 
$\bar{n}$ respectively, and $p_0^{\bar{\scriptscriptstyle \mathrm{D}}}$ is the
coalescence momentum of antideuteron. In the case where the momentum 
distributions of  $\bar p$ and $\bar n$ are isotropic and statistically independent, 
the energy spectrum of $\bar{\mathrm{D}}$ is
related to that of $\bar p$ and $\bar n$  as
\begin{align}\label{eq:p0-isotropic}
\gamma_{\bar{\scriptscriptstyle \mathrm{D}}}
\frac{d^{3}N_{\bar{\scriptscriptstyle \mathrm{D}}}}{d^{3}\vec{k}_{\bar{\scriptscriptstyle \mathrm{D}}}}
(\vec{k}_{\bar{\scriptscriptstyle \mathrm{D}}})
=
\frac{\pi}{6}
\left(p_{0}^{\bar{\scriptscriptstyle \mathrm{D}}}\right)^{3}
\cdot
\gamma_{\bar p} \frac{d^{3}N_{\bar p}}{d^{3}\vec{k}_{\bar p}}(\vec{k}_{\bar p})
\cdot
\gamma_{\bar n} \frac{d^{3}N_{\bar n}}{d^{3}\vec{k}_{\bar n}}(\vec{k}_{\bar n}) ,
\end{align}
where $\gamma_{\bar{\scriptscriptstyle \mathrm{D}}, \bar p, \bar n}$ are the 
Lorentz factors, and $\vec{k}_{\bar p}\approx \vec{k}_{\bar n}\approx 
\vec{k}_{\bar{\scriptscriptstyle \mathrm{D}}}/2$.

The coalescence criterion for the heavier anti-nuclei can be defined
in a similar way as that of antideuteron~\cite{Carlson:2014ssa,Cirelli:2014qia}.
For the case of $^{3}\overline{\mbox{He}}$, one can define the norms of the relative
four-momenta between the three anti-nucleons as three lengths
$ l_1 = ||k_1-k_2||,~~ l_2 = ||k_2-k_3||$ and $ l_3 =||k_1-k_3||$,  where $k_1, k_2, 
k_3$ are the four-momenta of the three anti-nucleons respectively. one can use 
these lengths to compose a triangle, and then make a circle  to envelope the 
triangle with a minimal diameter.  It is assumed that the three anti-nucleons can 
successfully merge into an anti-nucleus only if the diameter of the circle is less than
$p_0^{\scriptscriptstyle \overline{\mathrm{He}}}$~\cite{Carlson:2014ssa}.  
If the three lengths form a right or obtuse triangle (i.e.,
$l_i^2+l_j^2 \leqslant l_m^2$, for any $i, j, m$), the
minimal diameter equals to the longest side of the triangle, then the
criterion can be simply written as
$\text{max}\{l_1,l_2,l_3\} < p_0^{\scriptscriptstyle
  \overline{\mathrm{He}}}$.
On the other hand, if the three lengths form an acute triangle
($l_i^2+l_j^2 > l_m^2$, for all $i, j, m$), the minimal circle is just
the circumcircle of this triangle. In this case, the criterion can be expressed
in terms of the diameter of the circumcircle
\begin{equation}
\label{eq:pHebar-def1}
d_{\mathrm{circ}}=\frac{l_1l_2l_3}{\sqrt{(l_1+l_2+l_3)(-l_1+l_2+l_3)(l_1-l_2+l_3)(l_1+l_2-l_3)}}<
p_0^{\scriptscriptstyle \overline{\mathrm{He}}}.  
\end{equation}

An alternative way to define the coalescence criterion for
$^3\overline{\mathrm{He}}$ is simply requiring that the relative
four-momentum of each pair of the constituent anti-nuclei is smaller than
$\pHebar$~\cite{Cirelli:2014qia}:
\begin{equation} \label{eq:pHebar-def2}
||k_i-k_j|| < p_0^{\scriptscriptstyle \overline{\mathrm{He}}},
\ \  \ \ (i\neq j)  .
\end{equation}
If the relative four-momenta form a right or
obtuse triangle, This method is equivalent to the method of \eq{eq:pHebar-def1},
namely, $\pHebar$ is determined by the
longest side of the triangle.  However, for the case of acute triangles,
this method predicts slightly more anti-nuclei.  The quantitative difference
between these two methods will be discussed in the next section.

The spatial positions of particles also play an important role in
the formation of anti-nuclei, one should exclude the particle pairs
which are not close enough to each other in space.  As shown in
Ref. \cite{Carlson:2014ssa}, this can be taken into account by setting
all the particles with lifetime $\tau \gtrsim 2\,\, \mathrm{fm}/c$ to be
stable, where 2 fm is approximately the size of the  $^3\overline{\textrm{He}}$ nucleus.

\section{Coalescence momentum for antiHelium formation from the ALICE data}
\label{sec:ALICE}

The value of $p_0^{\bar{\mathrm{A}}}$ can  be constrained by collider data.
For instance, in the case of antideuteron production, the value of $\pDbar$ can be 
determined by reproducing the ALEPH measurement of the process 
$e^+e^-\to\overline{\textrm{D}}+X$ at the $Z^0$ resonance~\cite{Schael:2006fd}.  
An analysis based on the Monte-Carlo (MC) event generator  \texttt{PYTHIA} gave
$\pDbar= 0.192 \pm0.030~\textrm{GeV}$~\cite{Ibarra:2012cc}.
The value of $p_{0}^{\bar{\scriptscriptstyle\mathrm{D}}}$ is known to be slightly 
dependent on the center-of-mass (CM) energy and the collision 
process~\cite{Aramaki:2015pii,Adam:2015vda}.
As summarized in Ref. \cite{Aramaki:2015pii}, other  collider experiments at 
different center-of-mass (CM) energies lead to different values of 
$p_0$ in the range $\sim 
0.13-0.24$~GeV~\cite{Asner:2006pw,Lees:2014iub,Alper:1973my,Henning:1977mt,Sharma:2011ya,Chekanov:2007mv}.
Considering the resemblance to the dynamics of the annihilation of DM, we shall 
use  the value of $\pDbar$ derived from the ALEPH data as a benchmark value.

For the case of antihelium production, as the experimental data are rare,
it has been proposed to estimate the value of $\pHebar$ based on that of  
$\pDbar$. Two methods have been considered in literature~\cite{Carlson:2014ssa}.
The first one is to use the averaged ratio of $p^{A=3}_0/p^{A=2}_0$
through fitting the inclusive spectra of deuterons, tritons and $^3\mathrm{He}$
from the data on $AA$-collisions at the Berkeley Bevalac collider 
\cite{LEMAIRE197938}, and  then assume the relation
$\pHebar/\pDbar \approx \langle p_0^{A=3}/p_0^{A=2}\rangle$, which leads 
to
\beq{bevalac}
  p_{0}^{\scriptscriptstyle \overline{\mathrm{He}}} \approx
  \langle p_0^{A=3}/p_0^{A=2}\rangle 
  ~p_0^{\bar{\scriptscriptstyle \mathrm{D}}} =
  1.28~p_0^{\bar{\scriptscriptstyle \mathrm{D}}} = 0.246\pm0.038~\mathrm{GeV}.
  \eeq
The second one is to use the theoretical  scaling relation $p_0\sim 
\sqrt{E_b}$~\cite{Chardonnet:1997dv},
where $E_b$ is the total nuclear binding  energy, and obtain
\beq{binding}
  p_{0}^{\scriptscriptstyle \overline{\mathrm{He}}} 
  \approx
  p_0^{\bar{\scriptscriptstyle \mathrm{D}}}
  \sqrt{E_b^{^3\overline{\scriptscriptstyle \mathrm{He}}}/E_b^{\overline{\scriptscriptstyle \mathrm{D}}}}
   = 0.357\pm0.059~\mathrm{GeV}.
  \eeq

In recent years, direct production of antihelium/antitriton  have been observed in $pp$-collisions%
~\cite{Sharma:2011ya,Adam:2015vda,Acharya:2017fvb}  and $AA$-collisions at 
high  CM energies~\cite{Adler:2001uy,Agakishiev:2011ib}. These experiments 
measured the phenomenological coalescence parameter $B_A$  defined 
through the relation
\begin{align}
E_A\frac{d^3N_A}{dp_A^3}
=
B_A  \left(E_p\frac{d^3N_p}{dp_p^3}\right)^Z 
\left(E_n\frac{d^3N_n}{dp_n^3}\right)^N
, \ \ \ \vec{p}_p=\vec{p}_n=\vec{p}_A/A ,
\end{align}
where $A$ is the nucleus mass number, $Z$  the proton number and $N$ the 
neutron number with $A=Z+N$. In the isotropic limit,  it is expected that 
$B_A\approx p_0^{3(A-1)}$ from the phase-space analysis.
The ALICE experiment  measured the parameter $B_3$ for $\Hebar$ production 
in three transverse momentum bins $p_T/A=0.4$ -- 0.6~GeV, 0.6 -- 1.0~GeV and 
1.0 -- 2.0~GeV, respectively, with rapidity $|y|<0.5$ at a CM energy 
$\sqrt{s}=7$~TeV  in $pp$-collisions~\cite{Acharya:2017fvb}. 
The $B_3$ parameter for $\overline{\text{T}}$ production was measured in 
a single $p_T$ bin $p_T/A=0.4$ -- 0.6~GeV. The corresponding coalescence 
momenta were determined based on the values of $B_3$ using 
an interpolation  approach in which the relation $B_A\approx 
p_0^{3(A-1)}$ was assumed. 

As the real collision process could be  different  significantly from  the isotropic 
limit, we adopt an alternative approach in the determination of the coalescence 
momenta without assuming $B_A\approx p_0^{3(A-1)}$.
For a given MC event generator, we generate a large sample of 
$\mathcal{O}(10^{11})$ $pp$-collision events and keep record of the 
momentum information of the final states $\bar{p}/\bar{n}$  which have the 
potential to form an $\Hebar$ or $\overline{\text{T}}$ nucleus, namely, selecting 
the  $\bar{p}/\bar{n}$ particles according to a sufficiently large coalescence 
momentum $p_{0,\text{max}}=1$~GeV. We then allow the value of $p_0$ to vary freely in 
the range
$p_0 < p_{0,\text{max}}$ and use the condition of \eq{eq:pHebar-def1} for
antinucleus formation within the sample to fit the measured values of $B_3$.
We perform $\chi^2$-fits  for three MC event generators: 
\PYTHIA~\cite{Sjostrand:2006za,Sjostrand:2014zea}, 
\EPOSLHC~\cite{Werner:2005jf,Pierog:2013ria} ~and
\DPMJET~\cite{Roesler:2001mn}. 
Another popular event generator is {\tt QGS-JET}%
~\cite{Ostapchenko:2004ss,Ostapchenko:2005nj}.
It was shown in Ref. \cite{Kachelriess:2015wpa} that after some
tuning of the parameters for better fit the low-energy collider data,
the results from {\tt QGS-JET} is similar to that from \EPOSLHC.
The fit results  are summarized  in Tab~\ref{tab:p0_bestfit}. 
The uncertainties in the determined coalescence momenta arise from 
the uncertainties in the ALICE data, which are typically around $10\%$ or less.
Although our approach is quite different from that adopted by the ALICE 
collaboration, we find that both results are in reasonable 
agreement with each other.
\begin{table}[ht]
	\begin{tabular}{cccc}
		\hline\hline
		~MC generators: ~                       & ~~~~ \PYTHIA~~~~~  & ~~~~~\EPOSLHC ~~~~ & ~~~~\DPMJET ~~~~              \\ \hline
		$p_0^{\overline{\mathrm{He}}}$~(MeV) & $224^{+12}_{-16}~(254\pm14)$   & $227^{+11}_{-16}~(254\pm14)$     &  $212^{+10}_{-13}$           \\ \hline
		$p_0^{\bar{\mathrm{T}}}$~(MeV)    & $234^{+17}_{-29}~(266\pm22)$   & $245^{+17}_{-30}~(268\pm22)$     &  $222^{+16}_{-26}$           \\ \hline\hline
	\end{tabular}
	\caption{
	Best-fit values of $p_0^{\overline{\mathrm{He}}}$ and $p_0^{\bar{\mathrm{T}}}$ 
	from fitting to the ALICE data of $pp$-collision at $\sqrt{s}=7$~TeV for three MC 
	generators \PYTHIA~\cite{Sjostrand:2006za,Sjostrand:2014zea}, 
	\EPOSLHC~\cite{Werner:2005jf,Pierog:2013ria} and 
	\DPMJET~\cite{Roesler:2001mn}. The numbers in the brackets are the 
	values obtained by the ALICE collaboration using an interpolation 
	approach~\cite{Acharya:2017fvb}.
	}
	\label{tab:p0_bestfit}
\end{table}

\begin{figure}
	\includegraphics[width=13cm]{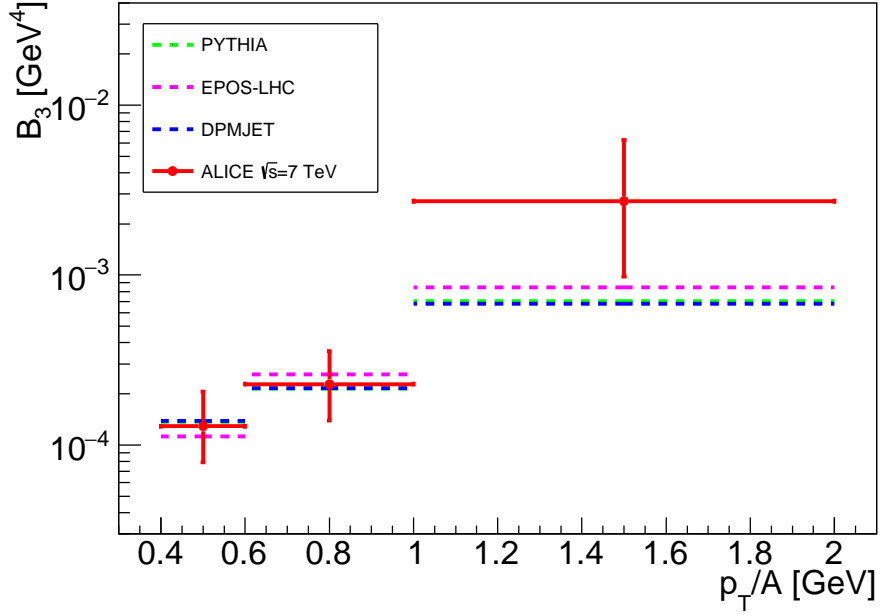} \caption{
	Best-fit values of the $B_3$ parameter from fitting to the ALICE
	data of $pp$-collision in three $p_T$ bins, for three MC
	generators \PYTHIA, \EPOSLHC~ and \DPMJET, respectively. The
	ALICE data are also shown~\cite{Acharya:2017fvb}.
	} \label{fig:b3_bestfit}
\end{figure}

\begin{figure}
	\includegraphics[width=7.3cm]{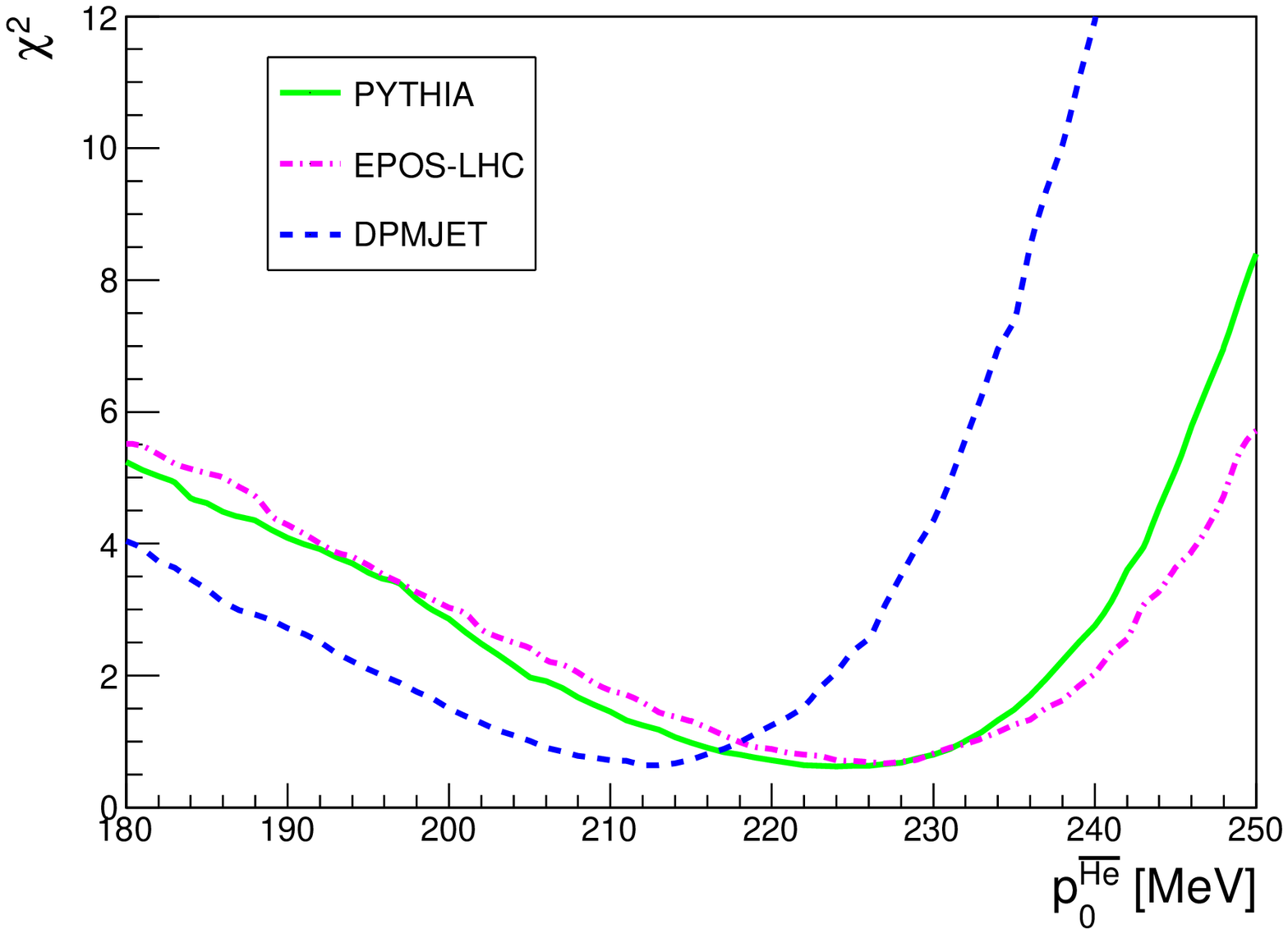}
        \includegraphics[width=7.3cm]{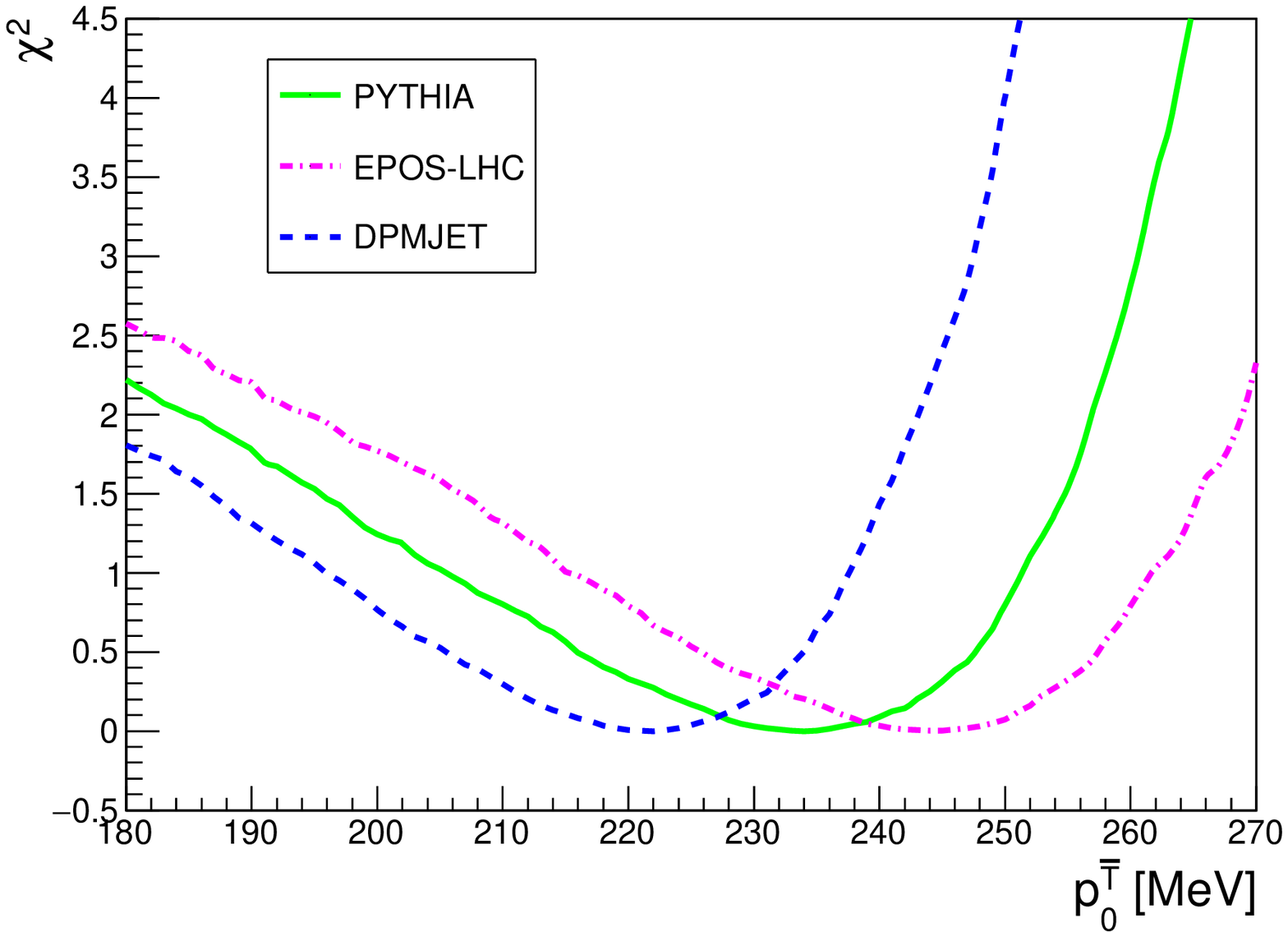} \caption{
	Left) Values of $\chi^2$ as a function of the coalescence
	momentum $\pHebar$ from fitting to the ALICE data
	~\cite{Acharya:2017fvb} for three MC event generators. Right)
	the same as left but for the coalescence momentum
	$p^{\overline{\text{T}}}_0$.}  \label{fig:chisq}
\end{figure}

In \fig{fig:b3_bestfit} we show the best-fit values of $B_3$ for
$\Hebar$ formation in three $p_T$ bins.
The $\chi^2$-curves of the fit results are shown in the left panel
of \fig{fig:chisq}. In all the fits, we find
$\chi^2_{\text{min}}/\text{d.o.f}\lesssim 0.6/2$, indicating
reasonable agreement with the ALICE data. The figure also shows that
the coalescence model can well reproduce the $p_T$-dependence of
$B_3$ in the low $p_T$ bins. At the highest $p_T$ bin
$p_T/A=1.0-2.0$~GeV, the coalescence model predict a slightly lower
value. For the case of $\overline{\text{T}}$ production, the ALICE
data is perfectly reproduced as it is only available for a single
$p_T$ bin. The corresponding $\chi^2$-curves are shown in the right
pannel of \fig{fig:chisq}.

An $\Hebar$ nucleus can be formed directly from the coalescence of  
$\bar{p}\bar{p}\bar{n}$, or through the $\beta$-decay of an antitriton 
$\overline{\text{T}}$ ($\bar{p}\bar{n}\bar{n}$).   Compared with the formation of 
$\overline{\text{T}}$, the direct formation channel is expected to be suppressed by 
Coulomb-repulsion between the two antiprotons. The suppression effect is, 
however, difficult to  estimate quantitatively. 
From \tab{tab:p0_bestfit}, it can  be seen that the determined  coalescence 
momenta for  $\Hebar$ are only slightly smaller than that for $\overline{\text{T}}$ 
by $\sim (5-10)\%$, suggesting that the effect of Coulomb-repulsion may not 
be significant, which is consistent with the analysis in Ref.~\cite{Blum:2017qnn}.
In the following calculations, we shall include the contributions from 
$\overline{\text{T}}$ using the corresponding  $B_3$ values, and 
approximate  the energy spectrum of $\Hebar$  from the decay of 
$\overline{\text{T}}$ to be the same as that of $\overline{\text{T}}$ 
for a given production process, which roughly enhance the final 
$\Hebar$ number by a factor of two or three depending on the production
processes.

\begin{figure}
\includegraphics[width=0.32\textwidth]{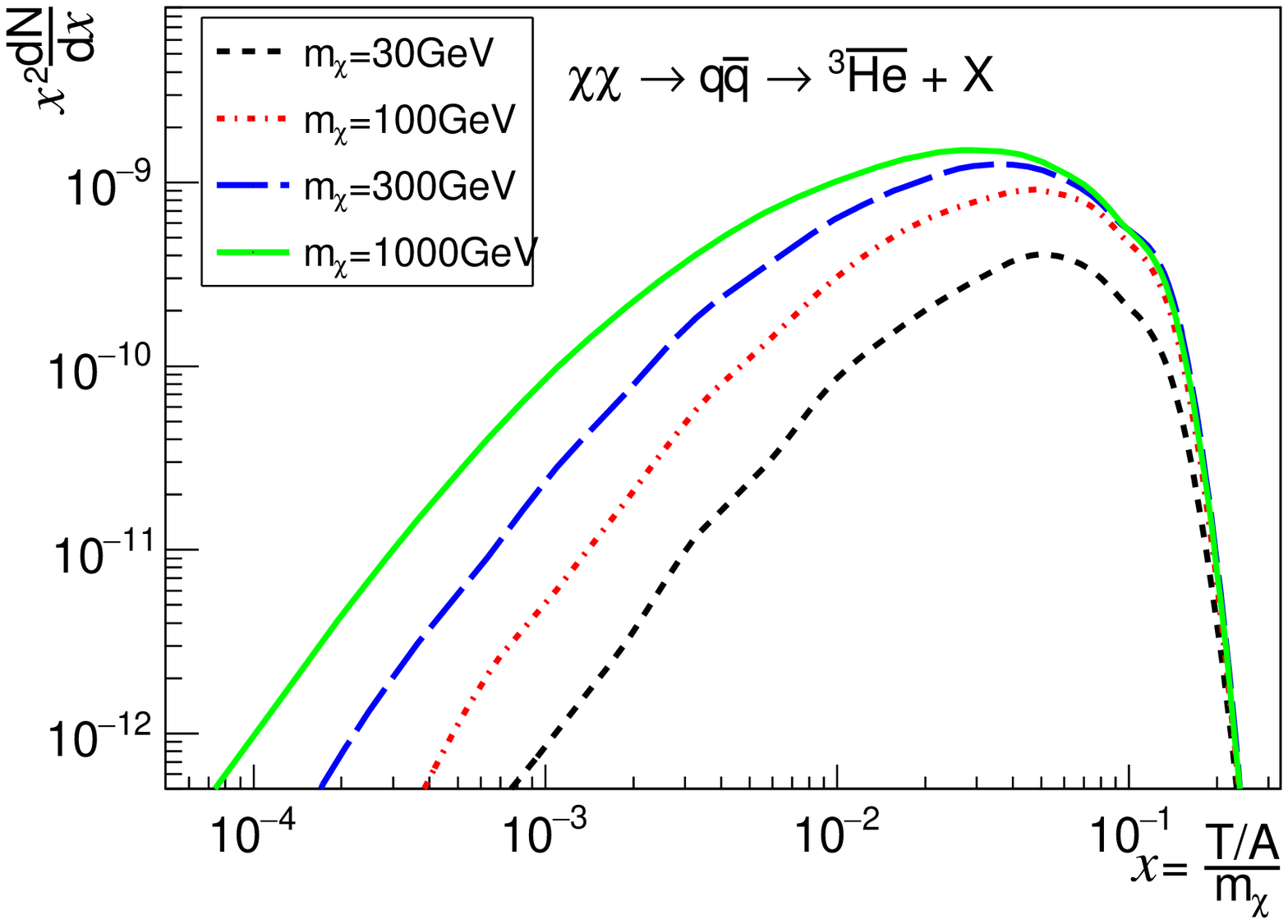}
\includegraphics[width=0.32\textwidth]{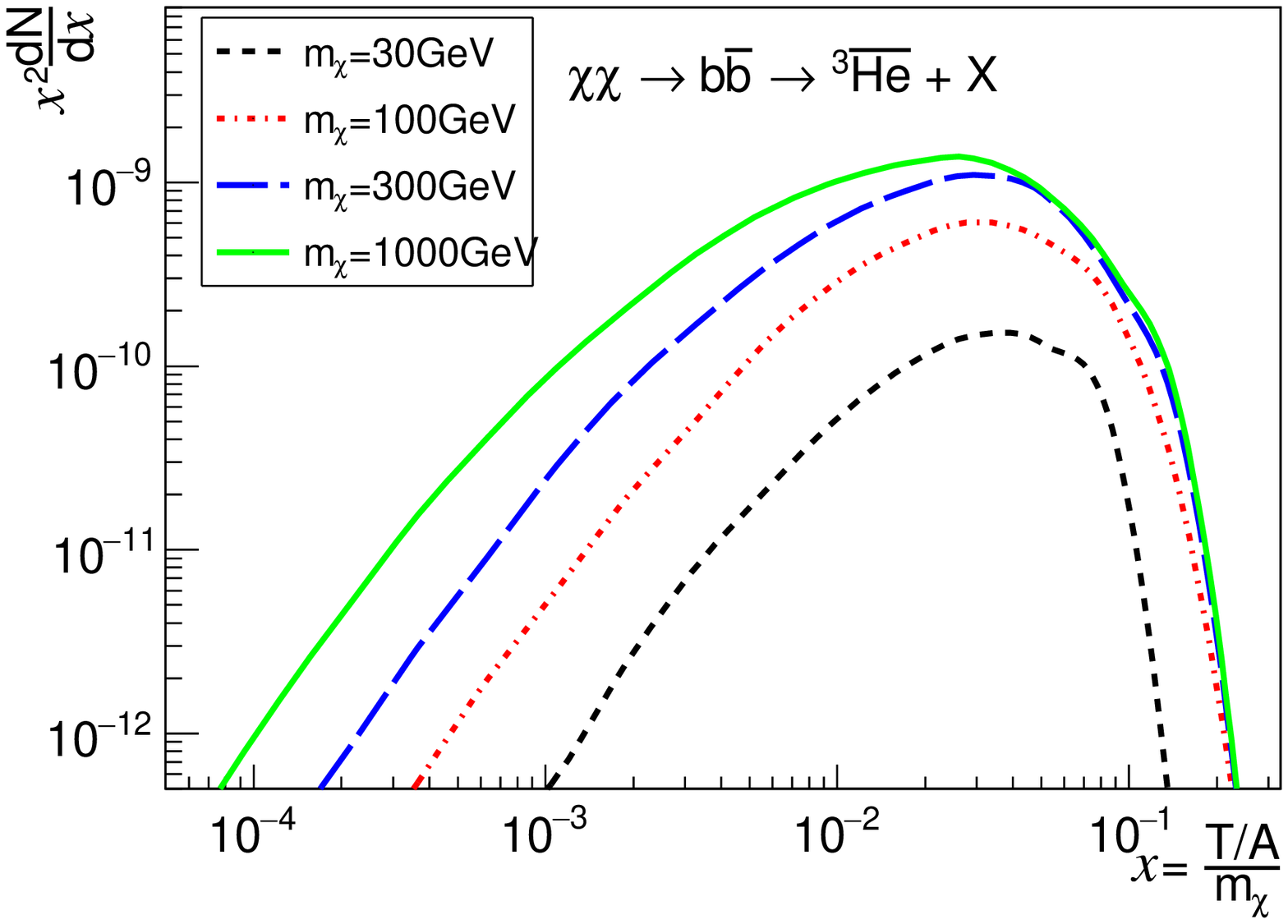}
\includegraphics[width=0.32\textwidth]{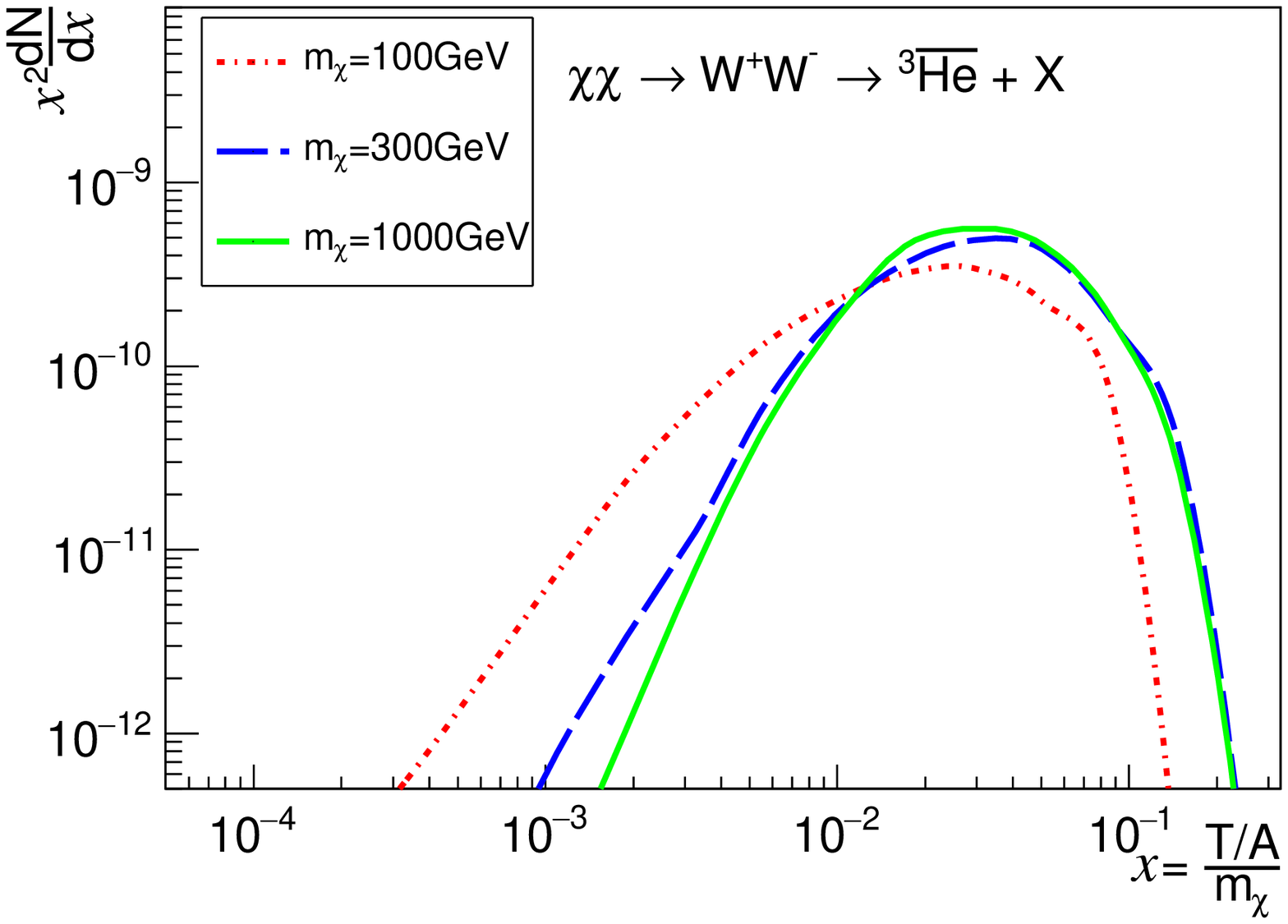}
\caption{
Energy spectra of CR $^3\overline{\textrm{He}}$ from DM annihilation
as a function of scaled kinetic energy per nucleon $x=T/(A m_{\chi})$
in the center-of-mass frame for different DM particle masses
$m_{\chi}=30-1000$~GeV.  The left, middle and right panels correspond
to $q\bar q$, $b \bar b$ and $W^{+}W^{-}$ annihilation channels,
respectively.  The events are generated using \PYTHIA
~\cite{Sjostrand:2006za,Sjostrand:2014zea}.
}
\label{spec}
\end{figure}

For the DM interaction induced primary $\Hebar$, we use \PYTHIA~
to simulate the hadronization processes of  DM annihilation, and adopt the 
coalescence model to describe the $\Hebar$ formation
on an event-by-event basis. 
In {\tt PYTHIA}, the DM annihilation process $\chi\chi\to f \bar f$ (where $f$  stands for  any SM particle) is mimicked  by a  process of electron-positron annihilation through a  fictious singlet scalar  $e^+ e^-\to \phi^*\to f \bar f$ with a center-of-mass energy  $\sqrt{s}=2 m_\chi$ and {\it all} the initial-state-radiations switched off, which guarantees that the initial states are color-neutral and not interfere  with the final state interactions of the SM particles. Other MC event generators, such as \EPOSLHC~ and {\tt DPMJET}, etc., only use hadrons as initial states, which are not appropriate to simulate the DM annihilation in a straightforward way.
We consider two  Majorana DM particles annihilating 
into $q\bar{q}$ ($q$ stands for $u$ or $d$ quark), $b\bar{b}$ and $W^+W^-$ final 
states.   Four possible values for the DM particle mass are considered,
$m_{\chi}\!\!= 30,\, 100,\, 300,$ and $1000$~GeV. We generate 
$\mathcal{O}(10^{11})$ events for each case of DM particle mass to produce 
enough 
$^3\overline{\textrm{He}}$ particles for calculating the injection energy spectrum.
The final number of $^3\overline{\textrm{He}}$ particles produced is
around $\mathcal{O}(10^{4})$ for all the cases.
In Fig. \ref{spec}, we show the obtained energy spectra of $\Hebar$ per DM 
annihilation as a function of the scaled kinetic energy per nucleon $x=T/(A 
m_{\chi}$).
As can be seen from \fig{spec}, for $q\bar{q}$ or $b\bar{b}$ channels, increasing 
the DM mass always leads to higher $\Hebar$ yields and softer spectrum due to 
the longer chain of parton showers. However, for $W^+W^-$  final states, the 
spectrum becomes harder. The reason is that for the heavier DM, the produced 
$W^+W^-$ particles  are more energetic, and the decay products of $W^+W^-$ 
are boosted to higher energies, which lead to the decrease of 
$^3\overline{\textrm{He}}$ towards lower energies.
In order to quantitatively discuss the difference between the two coalescence 
criterions in Eqs.~\eqref{eq:pHebar-def1} and \eqref{eq:pHebar-def2}, we perform 
a  test for the case of  $m_{\chi}=1000$ GeV with $q\bar{q}$ final states, and 
found that the method of Eq.~\eqref{eq:pHebar-def1} produces about $13\%$ 
fewer $^3\overline{\textrm{He}}$ particles compares to the method of 
Eq.~\eqref{eq:pHebar-def2}. The difference found in our calculation  is larger 
than the previous estimation of $\sim 6\%$ in Ref.~\cite{Carlson:2014ssa} which assumed an isotropic momenta distribution. Nevertheless, the difference between the two methods would not affect the conclusion qualitatively.

\begin{figure}
\includegraphics[width=0.48\textwidth]{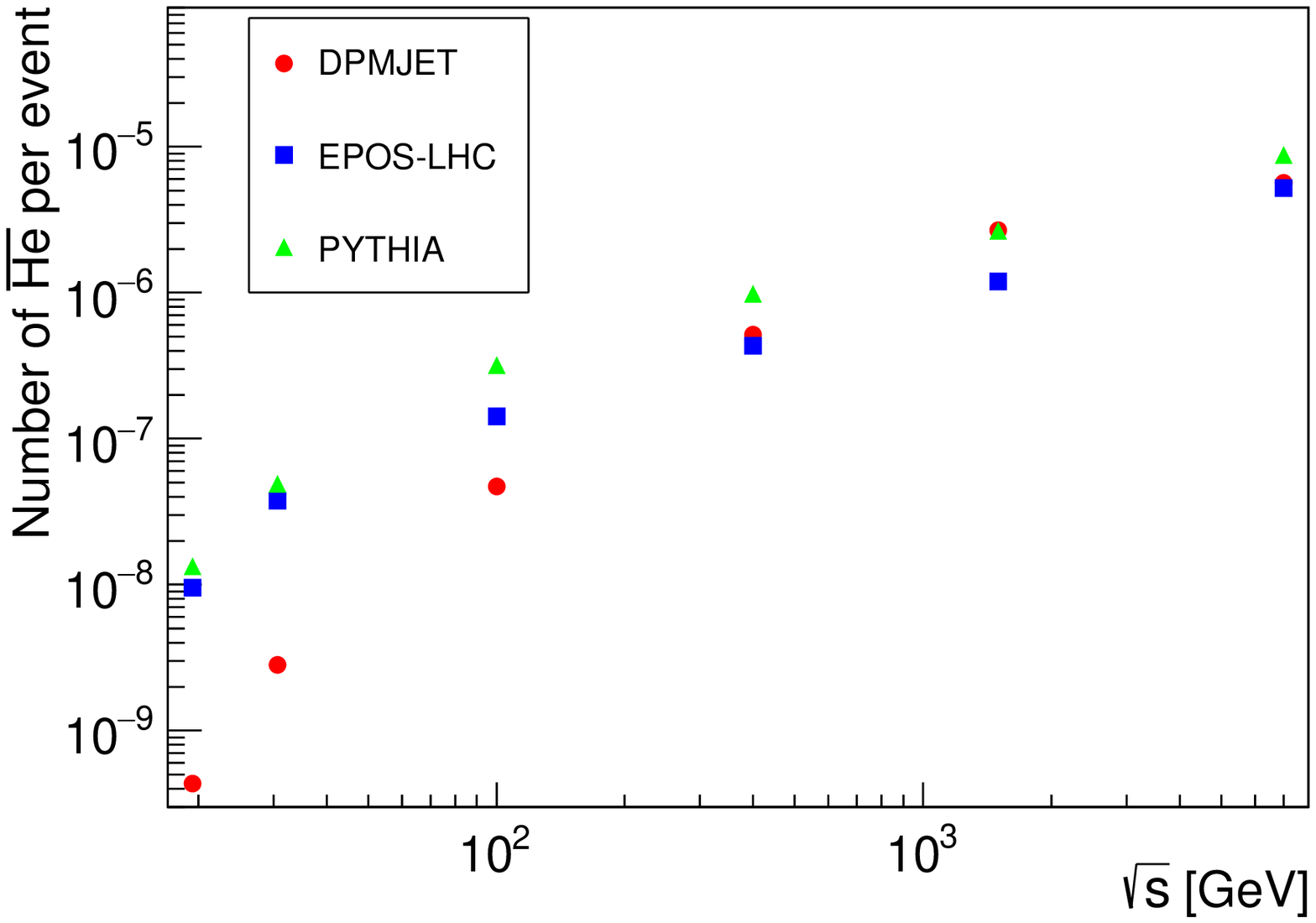}
\includegraphics[width=0.48\textwidth]{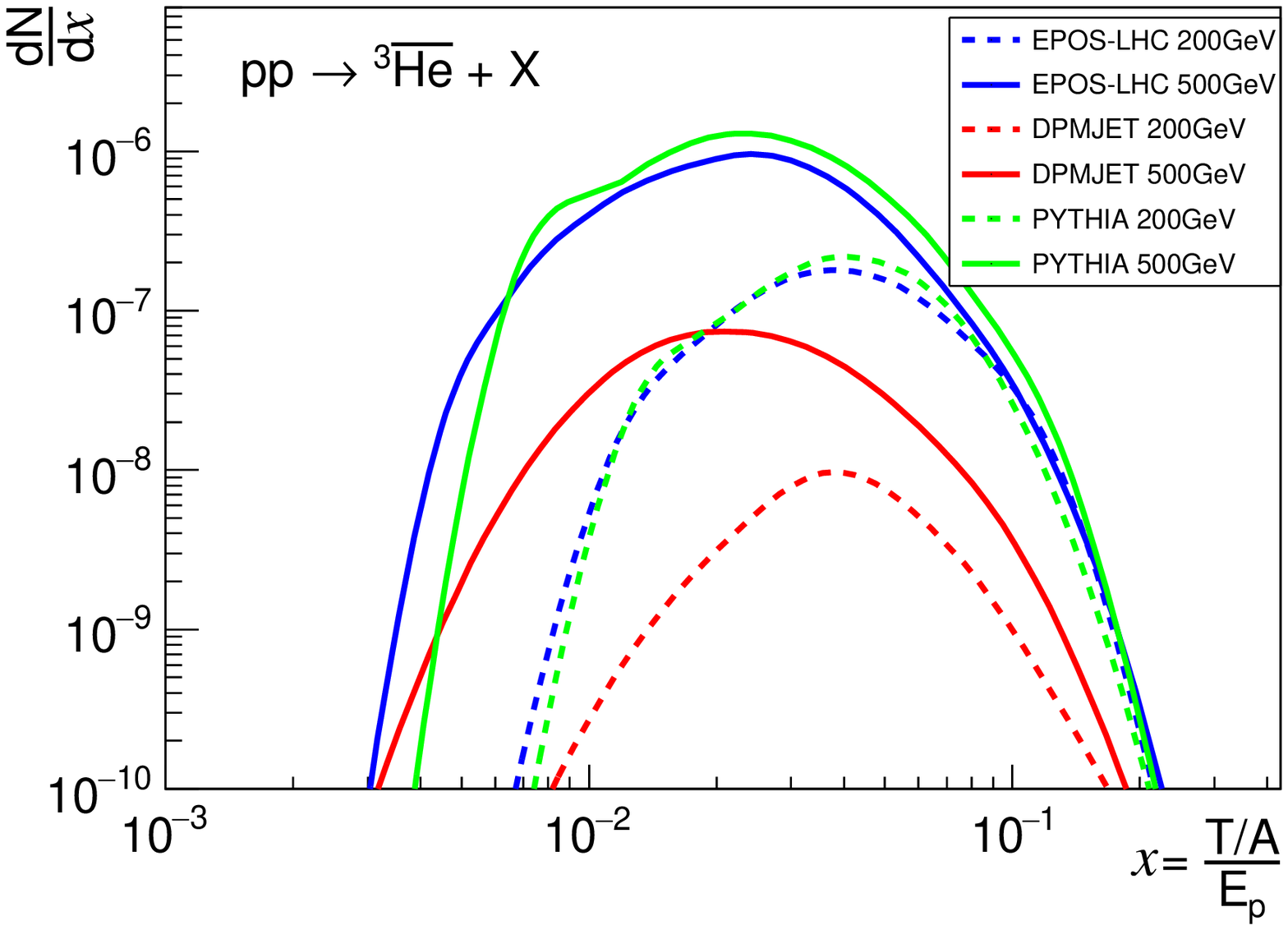}
\caption{
	Left) Total number of $\Hebar$ events from $pp$-collisions at different
	center-of-mass energies. The events are generated using three
	event generators \PYTHIA, \EPOSLHC~ and \DPMJET.  Right)
	Energy spectra of $^3\overline{\textrm{He}}$ events from
	$pp$-collisions in the target-rest frame at two different
	energies of incident protons $E_{\text{lab}}=200$ and 500 ~GeV
	for the three MC event generators.  }
\label{spec2}
\end{figure}

For the secondary $\Hebar$ productions from $pp$-collisions, we
shall use the MC event generators \EPOSLHC~ and \DPMJET~ to simulate
the production of anti-nucleons and the coalescence model to estimate
the formation of $\Hebar$. The difference between the two MC event
generators can be used as a rough estimation of the uncertainties
related to hadronization models. The default parameters in
\PYTHIA~(i.e. the Monash tune~\cite{Skands:2014pea}) is not optimized for $pp$-collisions at relatively low
energies. We anyway include the results from \PYTHIA~ for a comparison
purpose.
In the left panel of \fig{spec2}, we show the total number of $\Hebar$
events  at different CM energies for the three event generators.  
At CM energies around a few tens of
GeV which is most relevant to the secondary $\Hebar$
production, \PYTHIA~ and \EPOSLHC~give similar results
while \DPMJET~predict significantly lower number of $\Hebar$ events.
For instance, at $\sqrt{s}=30$~GeV, the difference between \EPOSLHC~
and \DPMJET~ can reach an order of magnitude.
The energy spectra obtained for the three event generators at two
different energies of incident protons $E_{\text{lab}}=200$ and
$500$~GeV (corresponds to $\sqrt{s}$=19.4 and 30.7~GeV, respectively) in 
the target rest frame are shown in the right panel of \fig{spec2}.
As it can be seen from the figure, although the coalescence momentum
of $\Hebar$ production in all the event generators are calibrated to
the same ALICE $pp$-collision data at $\sqrt{s}=7$~TeV, at lower CM
energies, the predicted $\Hebar$ energy spectrum can be significantly
different.

\section{Updated limits from AMS-02 antiproton data}\label{sec:propagation}
\subsection{Cosmic-ray propagation}
The propagation of CR anti-nuclei in the Galaxy can be described by a
diffusion model in which the diffusion zone is assumed to be a
cylinder with radius $r_h \approx 20$ kpc and half-height $z_h =
1 \sim 10$ kpc. The diffusion equation of CR charged particles can be
written as \cite{GINZBURG:1990SK,STRONG:2007NH}:
\beq{CR}
\frac{\partial f}{\partial t}= q(\vec{r},p)+\nabla\cdot(D_{xx}\nabla f-\vec{V}_c  f)+
\frac{\partial}{\partial p}p^2 D_{pp}\frac{\partial}{\partial p}\frac{f}{p^2}  - \frac{\partial}{\partial p}
\left[ \dot{p} f-\frac{p}{3}(\nabla\cdot\vec{V}_c) f \right] -\frac{f}{\tau_f}  - \frac{f}{\tau_r} ~,
\eeq
where $ f(\vec{r},p,t)$ is the number density per unit of particle
momentum $p$ at the position $\vec{r}$, and $q(\vec{r},p)$ is the
source term. $D_{xx}$ is the energy-dependent spatial diffusion
coefficient which is parameterized as $D_{xx} = \beta D_0
(R/R_0)^{\delta}$, where $R = p/(Ze)$ is the rigidity of the
cosmic-ray particle with electric charge $Ze$, $\delta$ is the
spectral power index which can take two different values $\delta
= \delta_{1(2)}$ when $R$ is below (above) a reference rigidity $R_0$,
$D_0$ is a constant normalization coefficient, and $\beta = v/c$ is
the velocity of CR particles. $\vec{V}_c$ is the convection velocity,
which is related to the galactic wind. Diffusive re-acceleration is
described as diffusion in momentum space, and is described by the
parameter $D_{pp}$ which can be parameterized as
 $D_{pp} = 4 V_a^2 p^2/(3 D_{xx} \delta(4-\delta^2)(4-\delta))$, 
 where $V_a$ is the Alfv\`{e}n velocity which characterises the 
 propagation of weak disturbances in a
magnetic field. $\dot{p} \equiv dp/dt$ is the momentum loss rate,
$\tau_f$ and $\tau_r$ are the time scales of particle fragmentation
and radioactive decay respectively.  The steady-state diffusion
condition is achieved by setting $\partial f/\partial t = 0$. For the
boundary conditions, it is assumed that the number densities of CR
particles are vanishing at the boundary of the halo: $ f(r_h,z,p)=
f(r,\pm z_h, p)= 0$.
We use the code \texttt{GALPROP
v54}~\cite{Strong:1998pw,MOSKALENKO:2001YA,%
STRONG:2001FU,MOSKALENKO:2002YX,PTUSKIN:2005AX} to solve the diffusion
equation of Eq.~\eqref{CR} numerically.

The source term in the propagation equation describe the creation of CR particles.
For the primary anti-nuclei $\bar{A}~ (\bar{A}=\bar{p}, \Hebar)$  produced by the 
annihilation of Majorana DM particles, the source term can be written as follows:
\beq{Hesource}
q_{\bar{A}}(\vec{r},p) = 
\frac{\rho_{_{\mathrm{DM}}}^2(\vec{r})}{2m^2_{\chi}}\langle\sigma v\rangle
\frac{dN_{\bar{A}}}{dp}~,
\eeq
where $\rho_{_{\mathrm{DM}}}(\vec{r})$ is the DM energy density, $\langle\sigma v\rangle$ is the
thermally-averaged annihilation cross section of DM and $dN_{\bar{A}}/dp$ is the 
energy spectrum of $\bar{A}$ discussed in the previous section.

The source term for the secondary $\bar{A}$ is given by

\beq{secsource}
q_{\bar A}(\vec{r},p) = 
\sum_{ij} n_j(\vec{r}) 
\int \beta_i \,c\, \sigma_{ij\to \bar A}^{\mathrm{inel}}(p^{\prime})
\frac{dN_{\bar{A}}(p,p^{\prime})}{d p}\,n_i(\vec{r},p^{\prime})\,dp^{\prime}~,
\eeq
where
$n_i$ is the number density of CR proton/Helium (or antiproton) per unit momentum, 
$n_j$ is the number density of the interstellar hydrogen/Helium, and
$\sigma_{ij}^{\mathrm{inel}}(p^{\prime})$ is the inelastic cross section for
the process $ij\to \bar A+X$. $dN_{\bar{A}}(p,p^{\prime})/d p$ is the
energy spectrum of $\bar{A}$ in the collision with the momentum of the incident
CR particle denoted by $p^{\prime}$. 
For the source term of $\bar p$, we include the collisions of $pp$, $p\text{He}$, 
$\text{He}p$, $\text{He}\text{He}$, $\bar p p$ and $\bar p \text{He}$.
For the source of $^3\overline{\mathrm{He}}$, since the $B_3$ data are
only available  in $pp$-collisions, we  consider the contribution form $pp$-collisions
which dominates the secondary background of $^3\overline{\mathrm{He}}$.

The energy spectrum of $\bar{p}$ can be obtained through some
parameterization formulae based on scaling behaviors with the involved
parameters determined by the low energy $pp$-collision data
~\cite{Tan:1983de,Duperray:2003bd,diMauro:2014zea,Korsmeier:2018gcy}, 
or using MC event generators~\cite{Kachelriess:2014mga,Kachelriess:2015wpa}. 
The difference between the two approaches can reach a factor of a 
few~\cite{Kachelriess:2015wpa}. The MC event generators can simulate 
the jet-structure of the final state partons which is very important for
the formation of heavy nuclei. In this work, for consistency, we shall use the 
MC event generators for both the secondary $\bar p$ and $\Hebar$ production
from $pp$-collisions.

The primary CR nucleus injection spectra are assumed to have a broken power 
law behavior $f_p(\vec{r},p)\propto p^{\gamma_p}$, with the injection index 
$\gamma_p = \gamma_{p1} (\gamma_{p2})$ for the nucleus rigidity $R_p$ below 
(above) a reference value $R_{ps}$. The spatial distribution of the interstellar gas 
and the primary sources of CR nuclei are taken from Ref.~\cite{Strong:1998pw}.
In the case of  $\bar{p}$ production, the tertiary contributions are included.
However, in the case of the $^3\overline{\textrm{He}}$ production,
the  sub-dominant tertiary contributions are neglected, as they are only compatible 
with the secondary background at kinetic energies below 
$0.4-0.6$~GeV/A~\cite{Korsmeier:2017xzj}. 

The inelastic interaction rate $\Gamma_{\mathrm{int}}$ of the
scattering between the nucleus $\bar{A}$ and the interstellar gas is
related to the fragmentation scale $\tau_f$ in Eq. \eqref{CR} as
$\Gamma_{\mathrm{int}}=1/\tau_f$, and can be estimated
as~\cite{Carlson:2014ssa,Cirelli:2014qia}
\beq{interact_rate}
\Gamma_{\mathrm{int}} = (n_{_{\mathrm{H}}} + 4^{2/3} 
n_{_{\mathrm{He}}})~v~\sigma_{\bar{A}p}~,
\eeq
where 
$n_{_{\mathrm{H}}}$ and $n_{_{\mathrm{He}}}$ are the number densities of 
interstellar hydrogen and helium, respectively, $4^{2/3}$ is the geometrical factor, 
$v$ is the velocity of $\bar{A}$ relative to interstellar gases, and 
$\sigma_{\overline{\textrm{He}}p}$ is the total inelastic cross section for 
the collisions between  $\bar{A}$ and the interstellar gas.
The number density  ratio $\mathrm{He/H}$ in the interstellar gas is taken to be
0.11~\cite{Strong:1998pw}.

Since the experimental data of the cross section
$\sigma_{\overline{\textrm{He}}p}$ is currently not available, we
assume the relation $\sigma_{\bar{A} p} = \sigma_{A \bar{p}}$ by
CP-invariance. For an incident nucleus with atomic mass number $A$, charge
number $Z$ and kinetic energy $T$, the total inelastic cross section
for $A\bar p$ collision is parameterized by the following
formula~\cite{MOSKALENKO:2001YA}
\beq{Apbar}
\sigma_{A\bar{p}}^{\mathrm{tot}} = A^{2/3}\left[48.2+19\, x^{-0.55} + (0.1-0.18\, x^{-1.2}) Z +
0.0012\,x^{-1.5}Z^2\right]~\mathrm{mb},
\eeq
where $x=T/(A \cdot \mathrm{GeV})$. For instance, by substituting
$A=3$ and $Z=2$, one obtains the cross section
$\sigma_{_{\overline{\textrm{He}} p}}$.

Finally, when anti-nuclei propagate into the heliosphere, the magnetic
fields of the solar system and the solar wind can distort the spectrum
of the charged CR particles. We use the force-field
approximation~\cite{Gleeson:1968zza} to quantify the effects of solar
modulation
\beq{force-field}
\Phi^{\mathrm{TOA}}_{A,Z}(T_{\mathrm{TOA}})=
\left(\frac{2m_A  T_{\mathrm{TOA}}+T^2_{\mathrm{TOA}}}{2m_A T_{\mathrm{IS}}+T^2_{\mathrm{IS}}}\right)
\Phi^{\mathrm{IS}}_{A,Z}(T_{\mathrm{IS}}),
\eeq
where $\Phi$ stands for the flux of the CR particles, which is related
to the density function $f$ by $\Phi = v f/(4\pi)$, ``TOA'' stands for
the value at the top of the atmosphere of the earth, ``IS'' stands for
the value at the boundary between the interstellar and the heliosphere
and $m$ is the mass of the nucleus.  $T_{\mathrm{IS}}$ is related to
$T_{\mathrm{TOA}}$ as $T_{\mathrm{IS}}=T_{\mathrm{TOA}}+e \phi_F|Z|$.
In this work, the value of the Fisk potential is fixed at $\phi_F=550$
MV.

\subsection{Updated upper limits on DM annihilation cross sections from 
AMS-02  antiproton data}\label{sec:pbarlimit}
The DM annihilation cross sections for annihilation channels such as
$\chi\chi\to q\bar{q}$, $b\bar{b}$ and $W^+W^-$ are subject to the constraints from
CR antiproton data, the constraints are expected to be strongly
correlated with the predictions for the maximal
$^3\overline{\textrm{He}}$ flux.
In Ref.~\cite{Jin:2015sqa}, upper limits on DM annihilation cross
sections were obtained based on the preliminary AMS-02 $\bar{p}/p$
data released in 2015~\cite{Ting:2015} and the background estimated
from the parametrization of Tan and Ng~\cite{Tan:1983de}. In this work,
we update the analysis by using the latest AMS-02
$\bar{p}/p$~\cite{Aguilar:2016kjl} and use MC event generators for calculating
antiproton production cross sections.

In order to take into account the uncertainties in CR propagation, we
consider three representative propagation models, the ``MIN'', ``MED''
and ``MAX'' models~\cite{JIN:2014ICA}. These models were obtained from
a global fit to the CR proton and B/C data of AMS-02 using
the \GALPROP~ code, which represents the typically minimal, median and
maximal antiproton fluxes due to the uncertainties in propagation
models. Note that these models are different from the ones proposed in
Ref.~\cite{Donato:2003xg} which are based on semi-analytical solutions
of the propagation equation. The values of parameters for the three
models are listed in Tab. \ref{tab2}.
In this updated analysis, the primary nuclei source term is normalized
to reproduce the AMS-02 proton flux at a reference kinetic energy $T =
100$ GeV, which is the default normalization scheme in {\tt GALPROP}.

\begin{figure}[ht]
	\includegraphics[width=14.5cm]{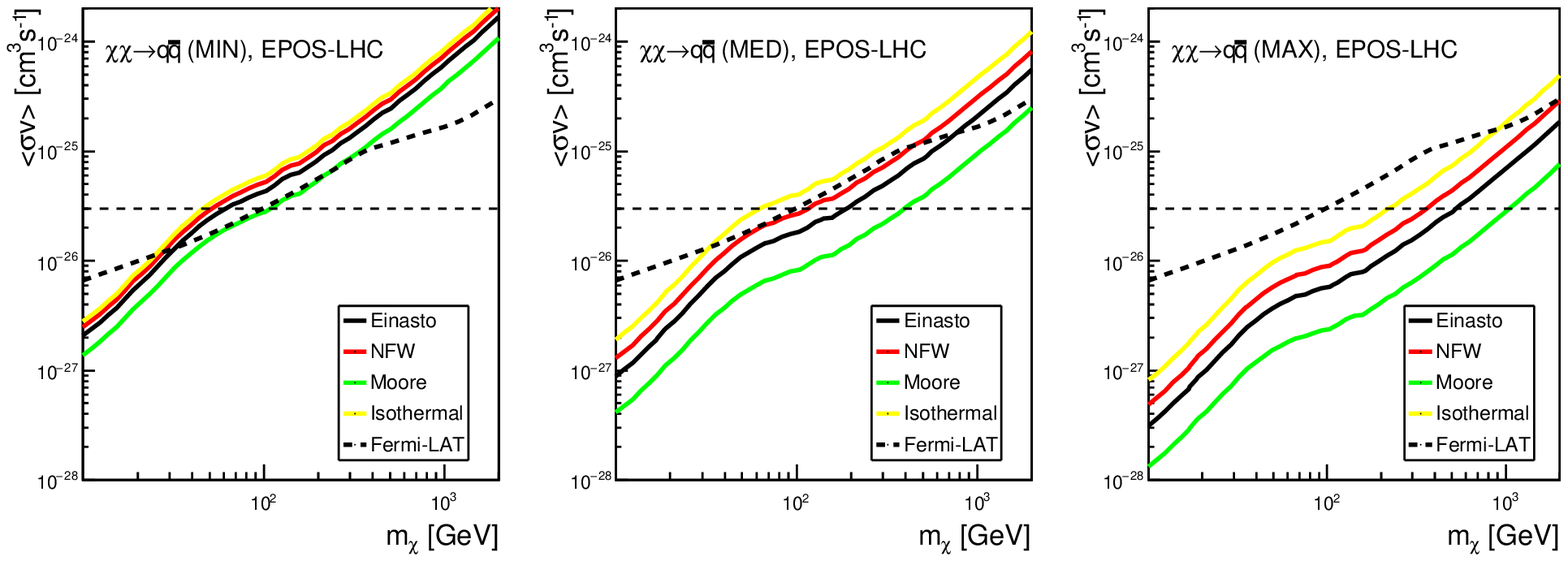}
	\includegraphics[width=14.5cm]{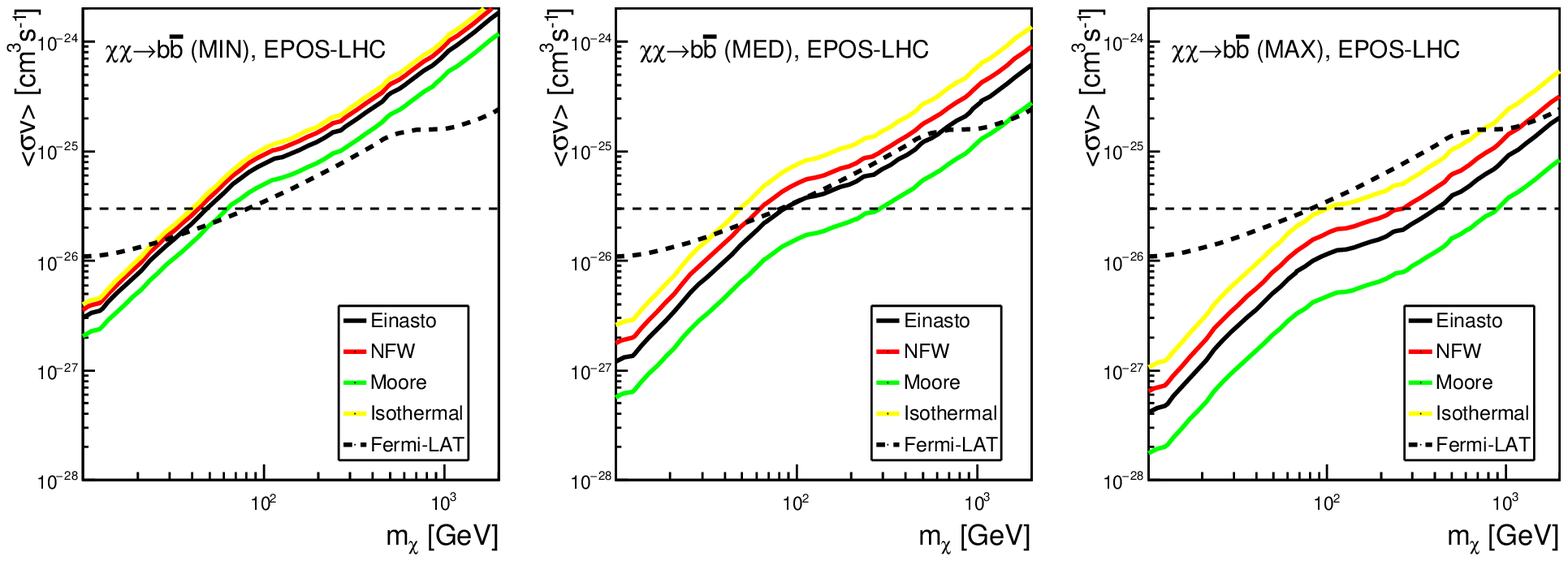}
	\includegraphics[width=14.5cm]{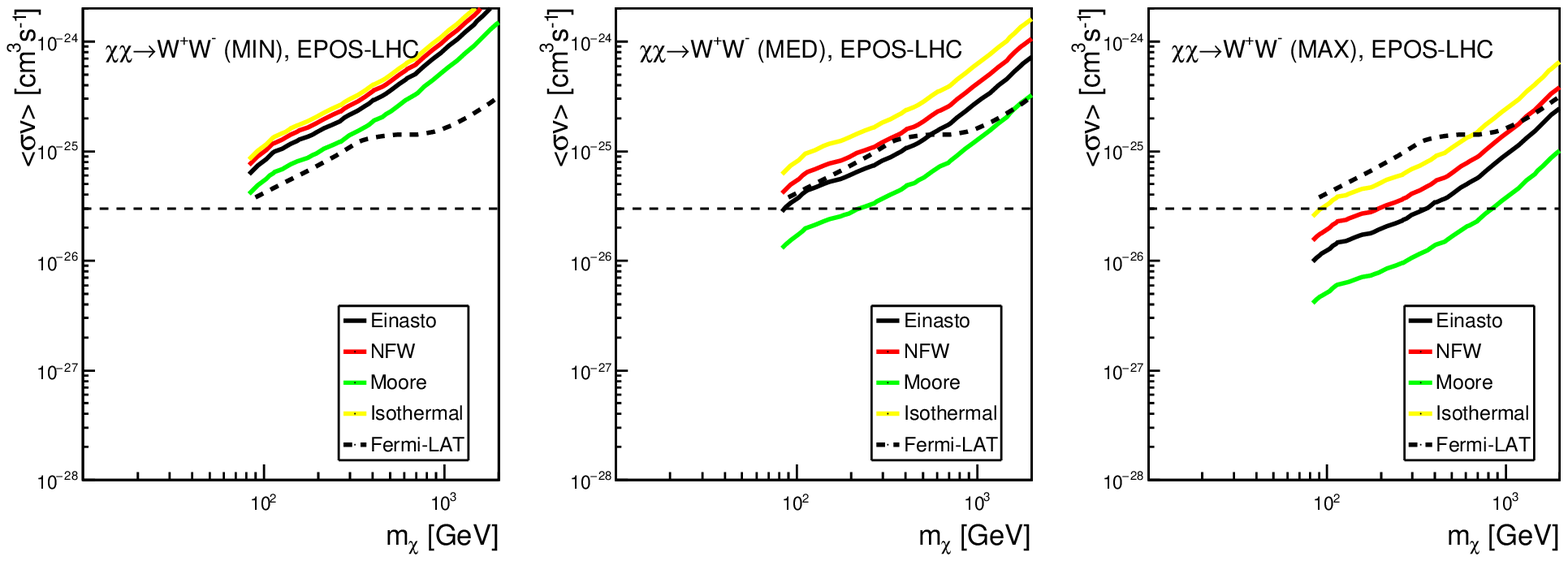}
	\caption{
	$95\%$ C.L. upper limits on  DM annihilation cross sections
	for different annihilation channels, propagation models and DM profiles.
	The secondary backgrounds are estimated using MC event generator \EPOSLHC.
	}
	\label{fig:limits-EPOS}
\end{figure}

\begin{figure}[ht]
	\includegraphics[width=14.5cm]{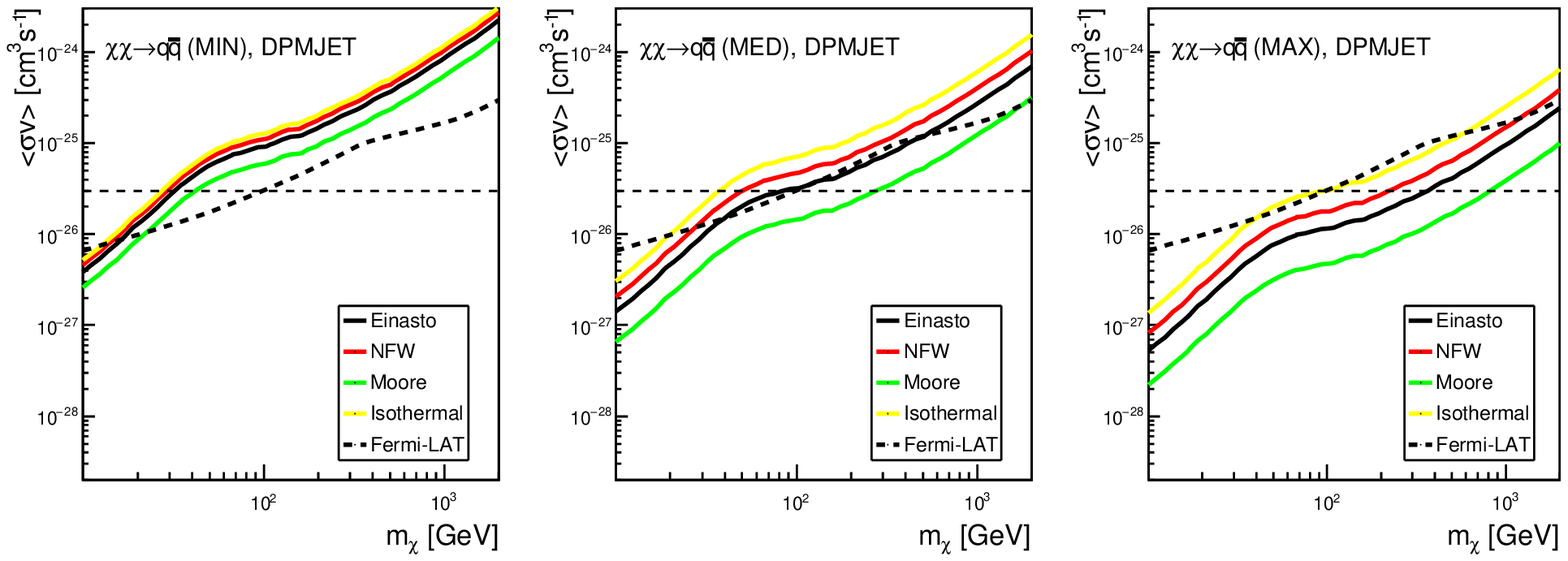}
	\includegraphics[width=14.5cm]{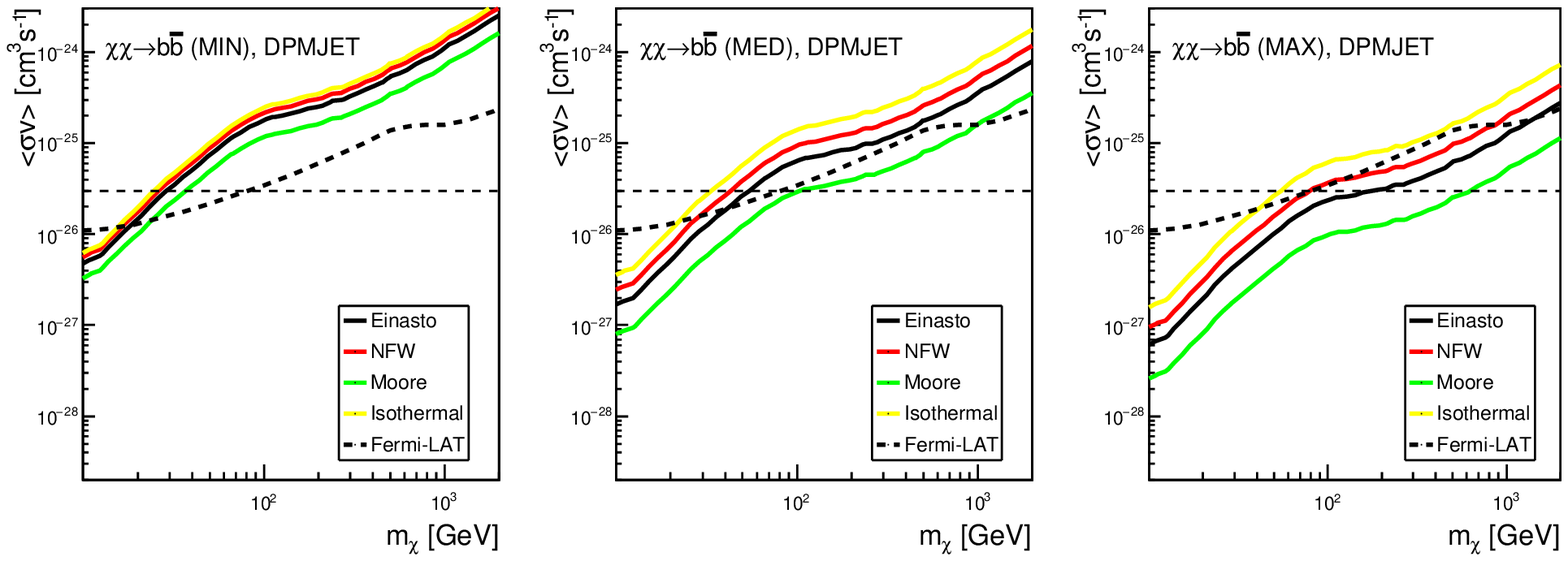}
	\includegraphics[width=14.5cm]{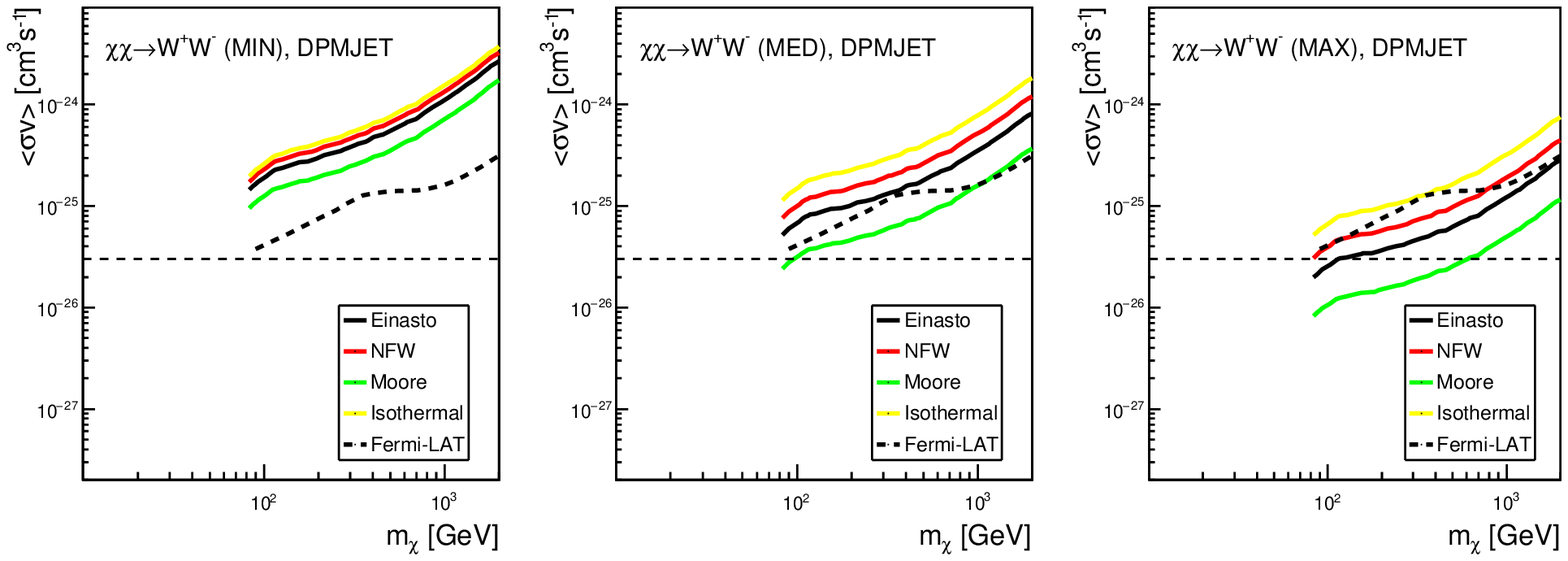}
	\caption{
	The same as Fig.~\ref{fig:limits-EPOS}, but with the secondary backgrounds estimated by \DPMJET.
	}
	\label{fig:limits-DPMJET}
\end{figure}

For each propagation model, four commonly used DM density profiles are
considered: 
the Navarfro-Frenk-White (NFW) profile \cite{NAVARRO:1996GJ}, 
the Isothermal profile \cite{Bergstrom:1997fj}, 
the Moore profile \cite{Moore:1999nt,Diemand:2004wh} and
the Einasto  profile \cite{Einasto:2009zd}.
The upper limits on the DM annihilation cross section as a function of DM particle mass are derived using the frequentist $\chi^{2}$-analyses. The expression of $\chi^2$ is defined as
$	\chi^2=\sum_i (f_i^{\text{th}}-f_i^{\text{exp}})^2/\sigma_i^2$,
	where $f_i^{\text{th}}$ are the theoretical predictions,
	$f_i^{\text{exp}}$ and $\sigma_i$ are the central values  and errors of experimental data, respectively. 
	The index $i$ runs over all the available data points.
	For a given DM particle mass, 
	we first calculate the minimal value $\chi_{\text{min}}^{2}$ of the $\chi^2$-function,
	and 
	then derive the  $95\%$ CL upper limits on the annihilation cross section,
	corresponding to  $\Delta \chi^{2}=3.84$ for one parameter.
More details of deriving the upper limits can be found in Ref.~\cite{Jin:2015sqa}. 
The obtained limits for $q\bar{q}$, $b\bar{b}$
and $W^+W^-$ channels are presented in Fig. \ref{fig:limits-EPOS} and
Fig.~\ref{fig:limits-DPMJET}, together with the secondary background estimated by  
\EPOSLHC~ and \DPMJET, respectively.

We find that the updated upper limits are comparable with the previous
ones, but the constraints for $m_{\chi} \lesssim 100$~GeV become more
stringent, which is partly due to larger cross sections from \EPOSLHC
~and \DPMJET~ and the updated AMS-02 data.
For a comparison, in Figs. \ref{fig:limits-EPOS}-\ref{fig:limits-DPMJET}, 
the upper limits from the Fermi-LAT 6-year gamma-ray data of the dwarf 
spheroidal galaxies \cite{Fermi-LAT:2016uux} are also shown. 
With the growth of the DM mass, the upper limits become weaker, and
can be well above the typical thermal relic cross section
$\langle\sigma v\rangle= 3\times 10^{-26}~ \mathrm{cm^3\cdot s^{-1}}$
for $m_\chi \gtrsim 100$~GeV. The upper limits for
$\chi\chi\rightarrow q\bar{q}$ channel are the most stringent among
the three types of final states.
For the ``MED'' and the ``MAX'' model, the obtained upper limits are
comparable with that from the Fermi-LAT $\gamma$-ray data.
\begin{table}[ht]
\begin{tabular}{ccccccccc}
\hline\hline
 Model         &~$r_h$(kpc)~&$~z_h$(kpc)~&~~$D_0$~~&~~$R_0$(GV)~~&~~~~~$\delta_1/\delta_2$~~~~~&$~V_a$(km/s)~&~$R_{ps}$(GV)~&~~~~$\gamma_{p1}/\gamma_{p2}$~~~~\\ \hline
 MIN           & 20       & 1.8      &  3.53   & 4.0   &  0.3/0.3          &  42.7     & 10.0   &  1.75/2.44               \\ \hline
 MED           & 20       & 3.2      &  6.50   & 4.0   &  0.29/0.29        &  44.8     & 10.0   &  1.79/2.45               \\ \hline
 MAX           & 20       & 6.0      &  10.6   & 4.0   &  0.29/0.29        &  43.4     & 10.0   &  1.81/2.46               \\ \hline\hline
\end{tabular}
\caption{
Values of the main parameters in
the  ``MIN'', ``MED'' and ``MAX'' models derived from fitting to the AMS-02 $B/C$ and
proton data based on the {\tt GALPROP} code~\cite{JIN:2014ICA}. The parameter $D_0$ is in units of  
$10^{28}~\mathrm{cm}^2\cdot\mathrm{s}^{-1}$.
}
\label{tab2}
\end{table}


\section{Prospects of detecting $^3\overline{\textrm{He}}$ events at 
AMS-02}\label{sec:prospects}

After obtaining the upper limits on the DM annihilation cross
sections, it is straight forward to discuss the prospects of detecting
$\Hebar$ in current and future experiments. The major source of
uncertainties in the prediction for $^3\overline{\textrm{He}}$ fluxes
involves:
\romannumeral 1) 
the uncertainties in the choice of DM density profiles. For a fixed DM
annihilation cross section, the variation in the choice of DM profiles
from the NFW profile to the isotheramal profile can lead to a change
in the final $\Hebar$ flux up to an order of magnitude.
\romannumeral 2) 
the uncertainties in the choice of different propagation models. As
shown in the previous works \cite{Carlson:2014ssa,Cirelli:2014qia}, 
it can change the predicted flux of $^3\overline{\textrm{He}}$ up to two orders of
magnitude for a given DM annihilation cross section.
\romannumeral 3) 
the uncertainties in the modeling of $\Hebar$ formation. As the
current data of $\Hebar$ production are only available at high CM
energies around TeV scale, the MC event generators calibrated
to the same high-energy data may give quite different 
predictions at low energies. From the left panel of \fig{spec2},
the differences between the MC event generators can reach an order 
of magnitude.
\romannumeral 4)
The uncertainties in the coalescence momentum
$p_0^{\scriptscriptstyle \overline{\mathrm{He}}}$. The production rate of
$^3\overline{\textrm{He}}$ is approximately proportional to
$(p_0^{\scriptscriptstyle \overline{\mathrm{He}}})^6$. Thus an
uncertainty of $\sim10\%$ in $p_0$ can be amplified to $\sim 60\%$ in
the predicted antihelium flux.

\begin{figure}
\includegraphics[width=0.42\textwidth]{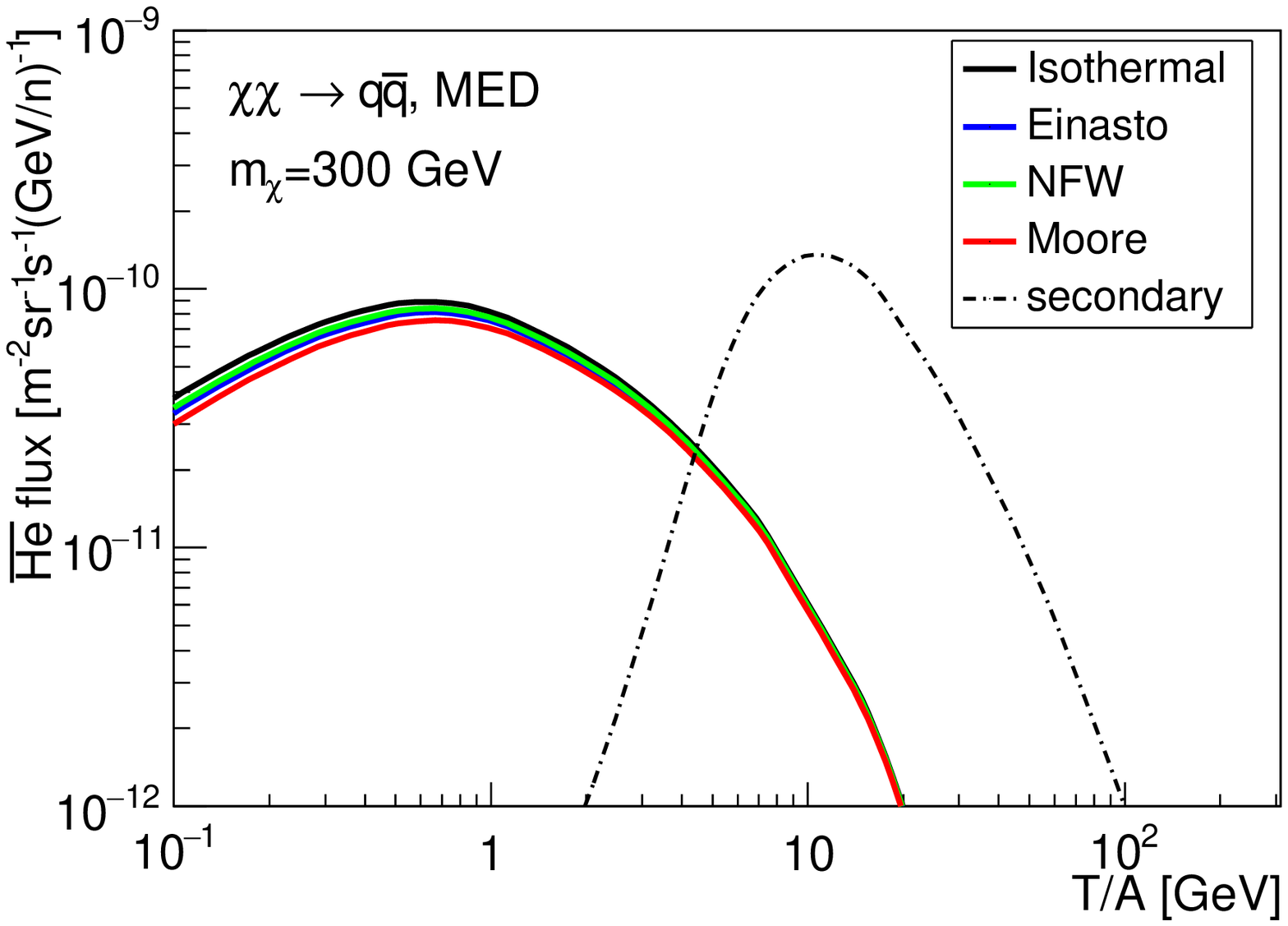}
\includegraphics[width=0.42\textwidth]{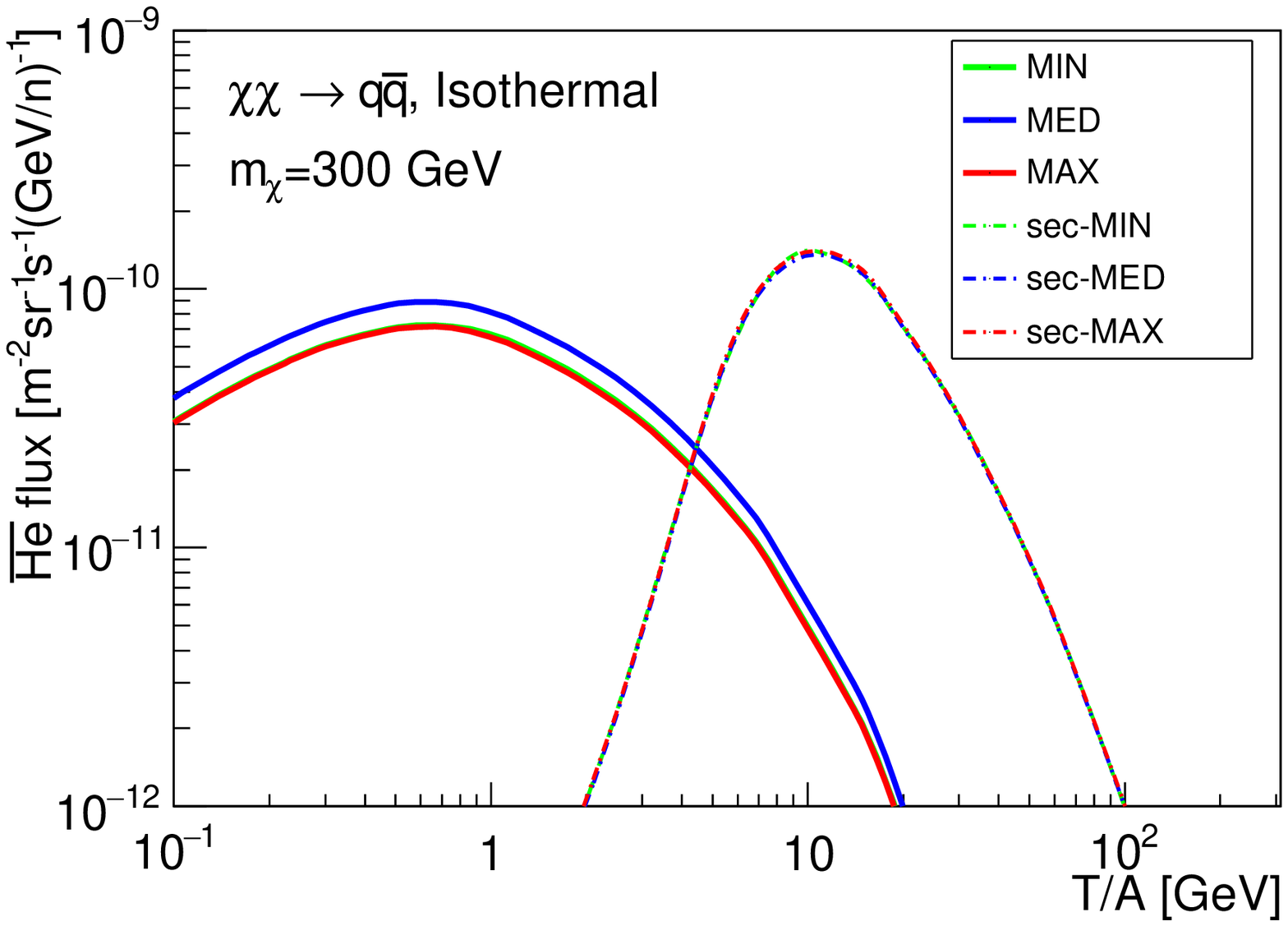}
\caption{
Left)
Predicted maximal $^3\overline{\textrm{He}}$ fluxes   (solid curves)
as a function of kinetic energy per nucleon
from DM annihilation into $q\bar q $ final states
in the MED propagation model
with four different DM profiles
NFW~\cite{NAVARRO:1996GJ},
Isothermal~\cite{Bergstrom:1997fj},
Einasto~\cite{Einasto:2009zd}
and  Moore~\cite{Moore:1999nt,Diemand:2004wh}.
The DM particle mass is fixed at $m_{\chi}=300$~GeV.
The secondary $^3\overline{\textrm{He}}$ fluxes generated 
by \EPOSLHC~ are also shown for a comparison.
Right) The same as left, but for three different propagation models
MIN, MED and MAX~\cite{JIN:2014ICA}
with DM profile fixed to  ``Isothermal''.
}
\label{fig4}
\end{figure}

An great advantage of using the antiproton data to constrain the predictions
for $^3\overline{\textrm{He}}$ flux is that the obtained constraints become highly
insensitive to the choice of DM density profile, as varying the DM
profile mainly results in a rescaling of the best-fit $\langle \sigma
v \rangle$ in such a way that the same antiproton flux is reproduced.
In the left panel of Fig.~\ref{fig4}, we show the prediction for the
maximal $^3\overline{\textrm{He}}$ flux after constrained by the
AMS-02 CR antiproton data for the four different DM profiles in the
same MED propagation model for DM particles with mass fixed at
$300$~GeV and $q\bar q$ the dominant annihilation final states. 
Compared with \fig{fig:limits-EPOS} and \fig{fig:limits-DPMJET},
it can be seen that  for the four DM profiles,
the difference in the  constraints on the DM annihilation cross sections 
can reach $\mathcal{O}(10)$, while that in the predicted 
$^3\overline{\textrm{He}}$ flux are reduced to  $\sim 30\%$.
Similarly, the predictions become also highly insensitive to the
choice of propagation models, provided that they give rise to similar
secondary antiproton backgrounds.  In the right panel of
Fig.~\ref{fig4}, we show the upper limits on
$^3\overline{\textrm{He}}$ flux for the three different propagation
models with the same DM profile. For the models giving nearly the same
secondary background such as the ``MIN'', ``MED'' and ``MAX'' models,
the difference in the predicted $^3\overline{\textrm{He}}$ flux 
is also very small $\sim 30\%$.

Detecting CR antihelium is one of the major scientific objectives of
the AMS-02 experiment. In this work, we give an estimation of the
maximal number of $^3\overline{\textrm{He}}$ events which can be
observed by AMS-02 after taking into account the constraints from the
AMS-02 antiproton data. We assume the whole lifetime of the experiment
to be 18 years, and adopt the most optimistic assumptions related to
the detectors.
In general, the number of $^3\overline{\textrm{He}}$ events observed
by a detector can be written as
\beq{count}
N=\int^{T_{\text{max}}}_{T_{\text{min}}} \eta~\Phi_{\scriptscriptstyle \overline{\mathrm{He}}} ~\mathcal{A} ~t ~d\, T ,
\eeq
where $\Phi_{\scriptscriptstyle \overline{\mathrm{He}}}$ is the flux
of $^3\overline{\textrm{He}}$, $\mathcal{A}$ is the acceptance of
$^3\overline{\textrm{He}}$ which is assumed to be identical to the
geometric acceptance of the AMS-02 detector $\mathcal{A}\approx
0.5~ \mathrm{m^2}\cdot\mathrm{sr}$, $\eta$ is the detecting efficiency
which is assumed to be unity, and $t \approx 18$~yr is the total
exposure time of the AMS-02 experiment. The lower and upper limits of
the integration are set to be $T_{\text{min(max)}}/A=0.1$~GeV (1 TeV).

In Tab.~\ref{tab:antihelium-num}, we show the predicted maximal number of
antihelium events for the three different annihilation channels and
four DM particle masses from 30~GeV to 1~TeV using two MC event
generators {\tt EPOS-LHC} and {\tt DPMJET}.  The expected secondary backgrounds
are also shown for a comparison.
We find that in the most optimistic cases, the expected total number of 
events can reach $\mathcal{O}(1)$, and is very likely to be dominated by
the secondaries.

\begin{table}
	\begin{tabular}{ccccccccc}
		\hline\hline
		
		~&$m_{\chi}$ (GeV)&~~~~~~~$\chi\chi\rightarrow q\bar{q}$~~~~~~~&~~~~~~~$\chi\chi\rightarrow b\bar{b}$~~~~~~~&~~~$\chi\chi\rightarrow W^+W^-$~~~   \\ \hline
		\multirow{4}*{$\mathrm{DM}$}
		&~30    ~&   ~$0.084 ^{+0.038}_{-0.040}$ $(0.153 ^{+0.070}_{-0.073})$~  &   ~$0.041 ^{+0.020}_{-0.018}$ $(0.073 ^{+0.036}_{-0.032})~$  &  ~ ---   ~                  \\
		&~100     ~&  ~$0.153 ^{+0.065}_{-0.072}$ $(0.269 ^{+0.114}_{-0.127})$~ &    ~$0.227 ^{+0.107}_{-0.103}$ $(0.419 ^{+0.198}_{-0.190})$~  &   ~$0.164 ^{+0.077}_{-0.076}$ $(0.304 ^{+0.143}_{-0.141})$~           \\
		&~300     ~&  ~$0.122 ^{+0.055}_{-0.056}$ $(0.179 ^{+0.081}_{-0.082})$~  &   ~$0.160 ^{+0.074}_{-0.074}$ $(0.256 ^{+0.118}_{-0.118})$~  & ~$0.054 ^{+0.025}_{-0.025}$  $(0.084 ^{+0.039}_{-0.039})$~         \\
		&~1000    ~&   ~$0.106 ^{+0.048}_{-0.048}$ $(0.138 ^{+0.063}_{-0.063})$~ & ~$0.131 ^{+0.058}_{-0.061}$  $(0.179 ^{+0.079}_{-0.083})$~ &  ~$0.015 ^{+0.007}_{-0.007}$ $(0.019 ^{+0.009}_{-0.009})$~         \\ \hline
		\multirow{2}*{~Secondary~} & \multicolumn{4}{c}{\multirow{2}*{$0.986 ^{+0.437}_{-0.455}$~$(0.054 ^{+0.021}_{-0.021})$}}\\
		\\ \hline\hline
	\end{tabular}
	\caption{
	Prospective maximal number of $^3\overline{\textrm{He}}$ particles
	with which can be detected by AMS-02 after 18 years of data taking under the most 
	optimistic assumptions. 
	The number of secondary $\Hebar$ 
	are estimated using MC event generator  \EPOSLHC. The numbers in the brackets
	correspond to the results  using \DPMJET. The quoted uncertainties  are due to  that in 
	the coalescence momentum $\pHebar$ determined from the ALICE data.}
	\label{tab:antihelium-num}
\end{table}

\begin{figure}[ht]	
	\includegraphics[width=0.32\textwidth]{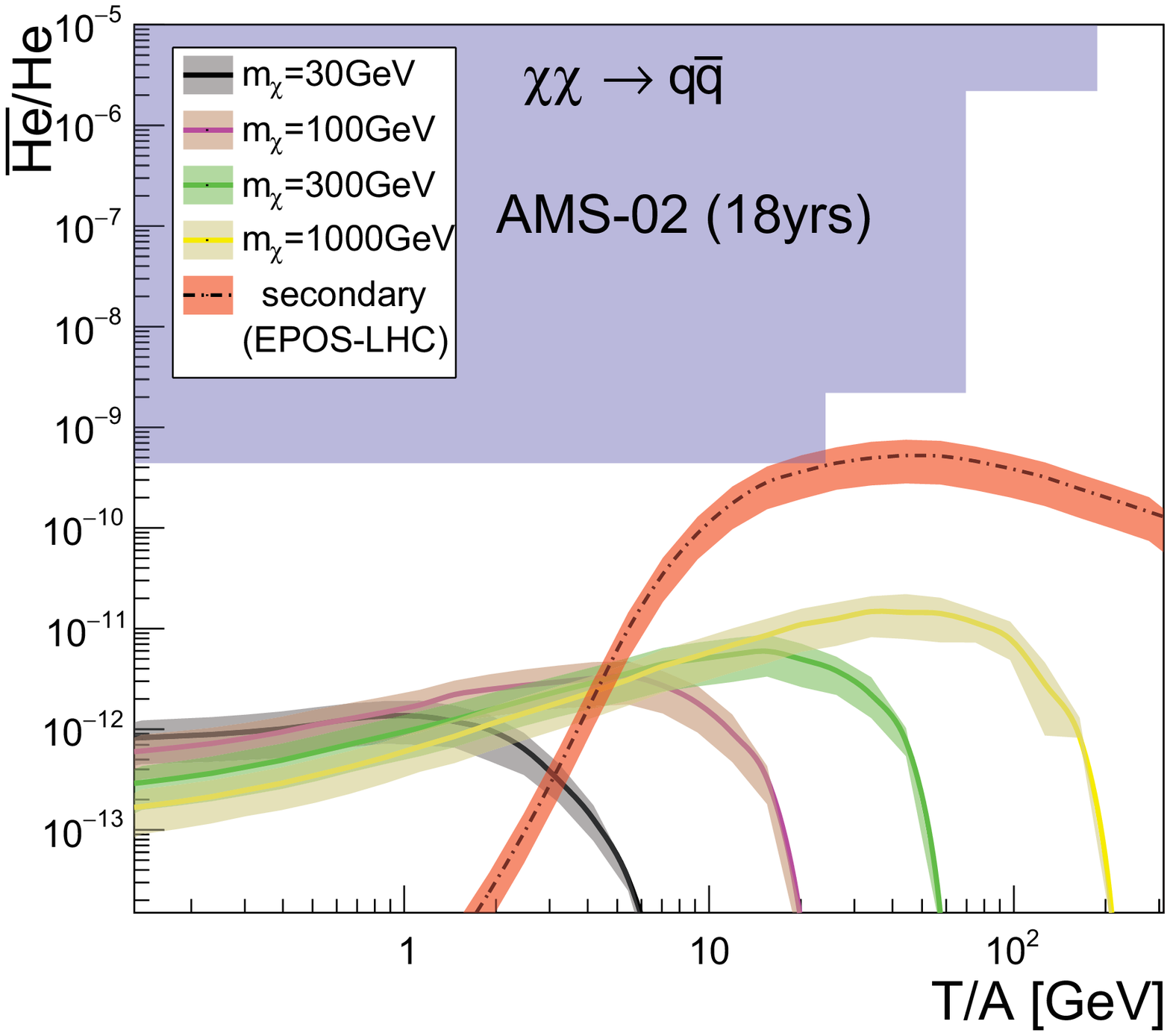}
	\includegraphics[width=0.32\textwidth]{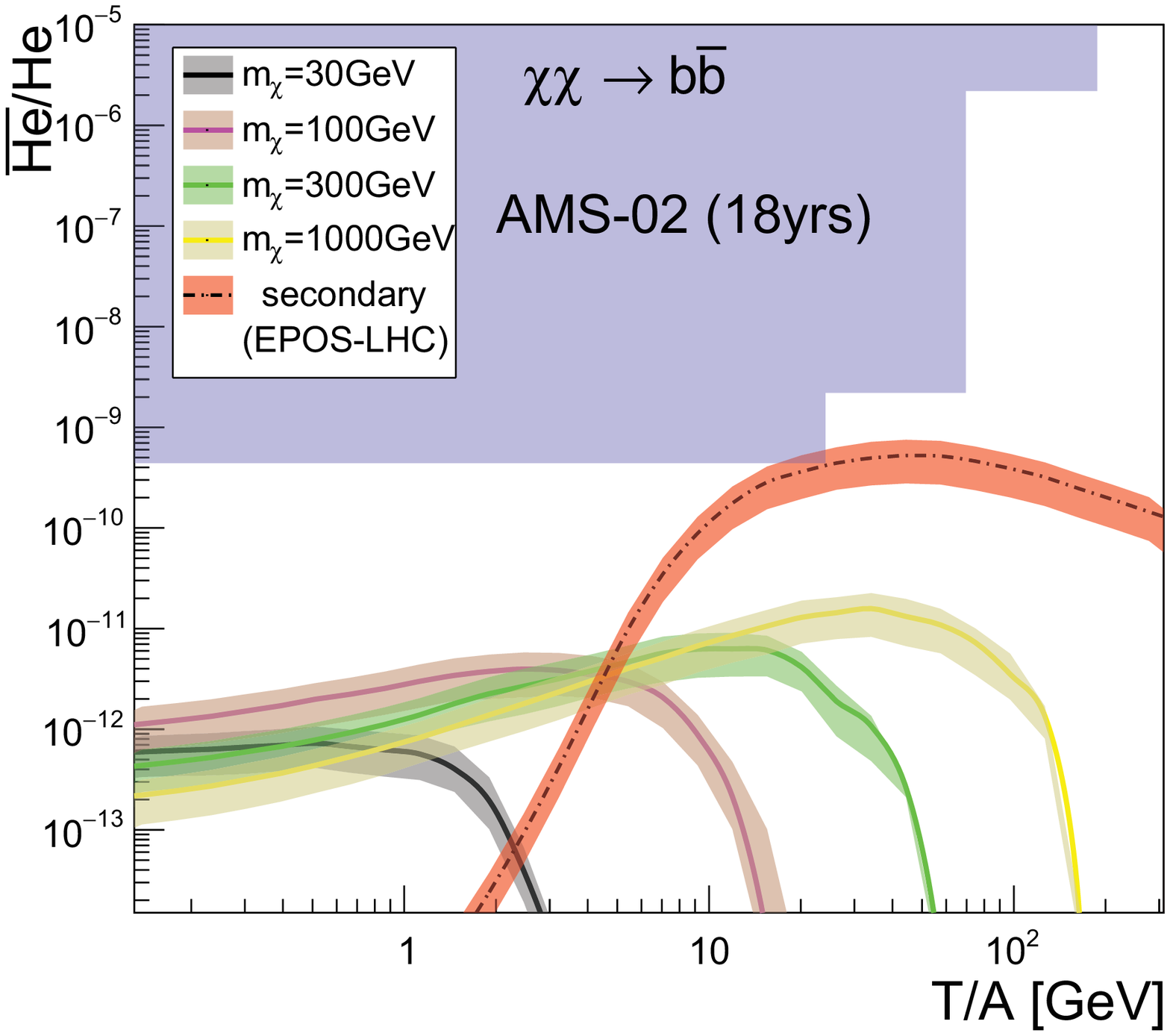}
	\includegraphics[width=0.32\textwidth]{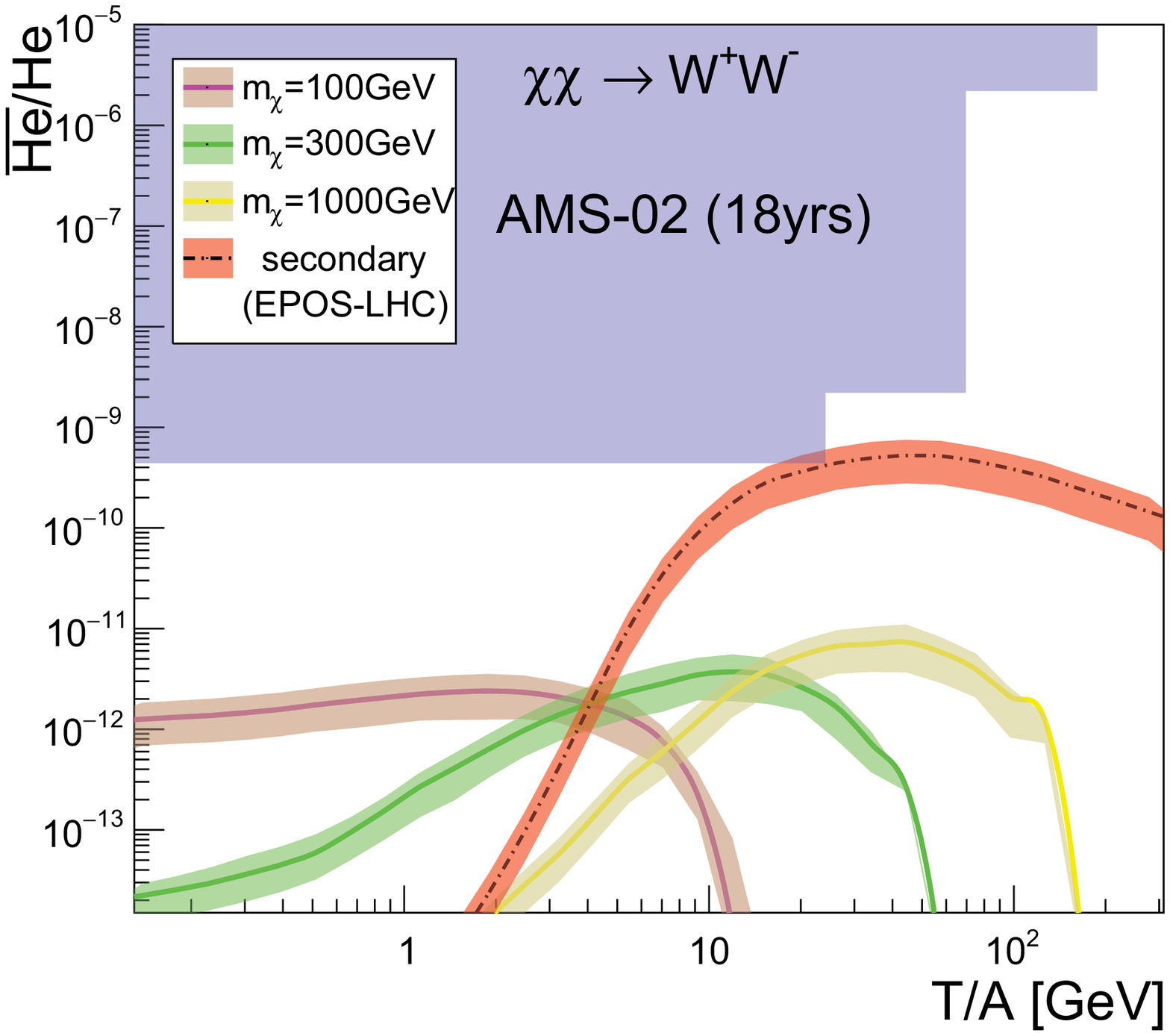}\\
	\includegraphics[width=0.32\textwidth]{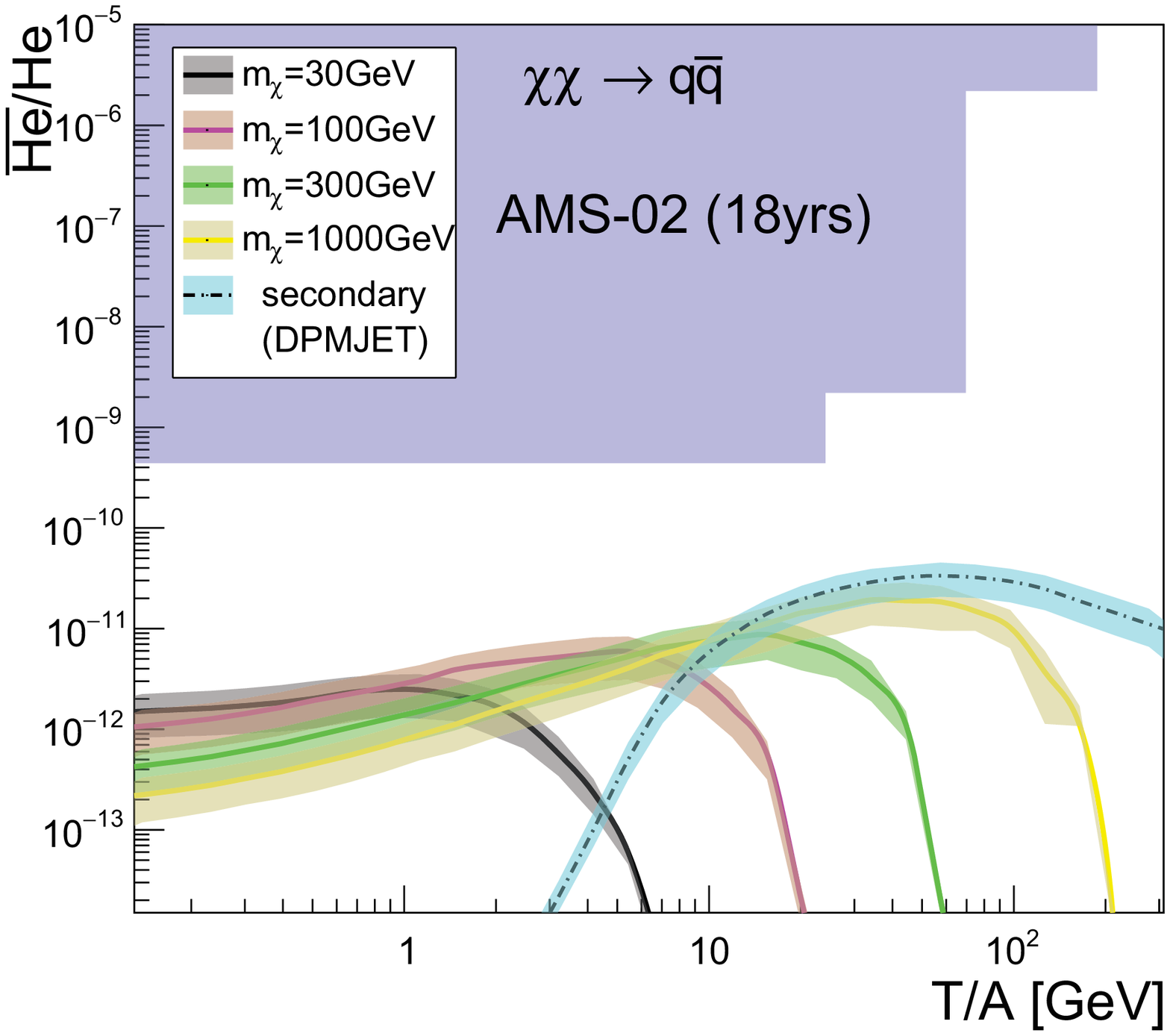}
	\includegraphics[width=0.32\textwidth]{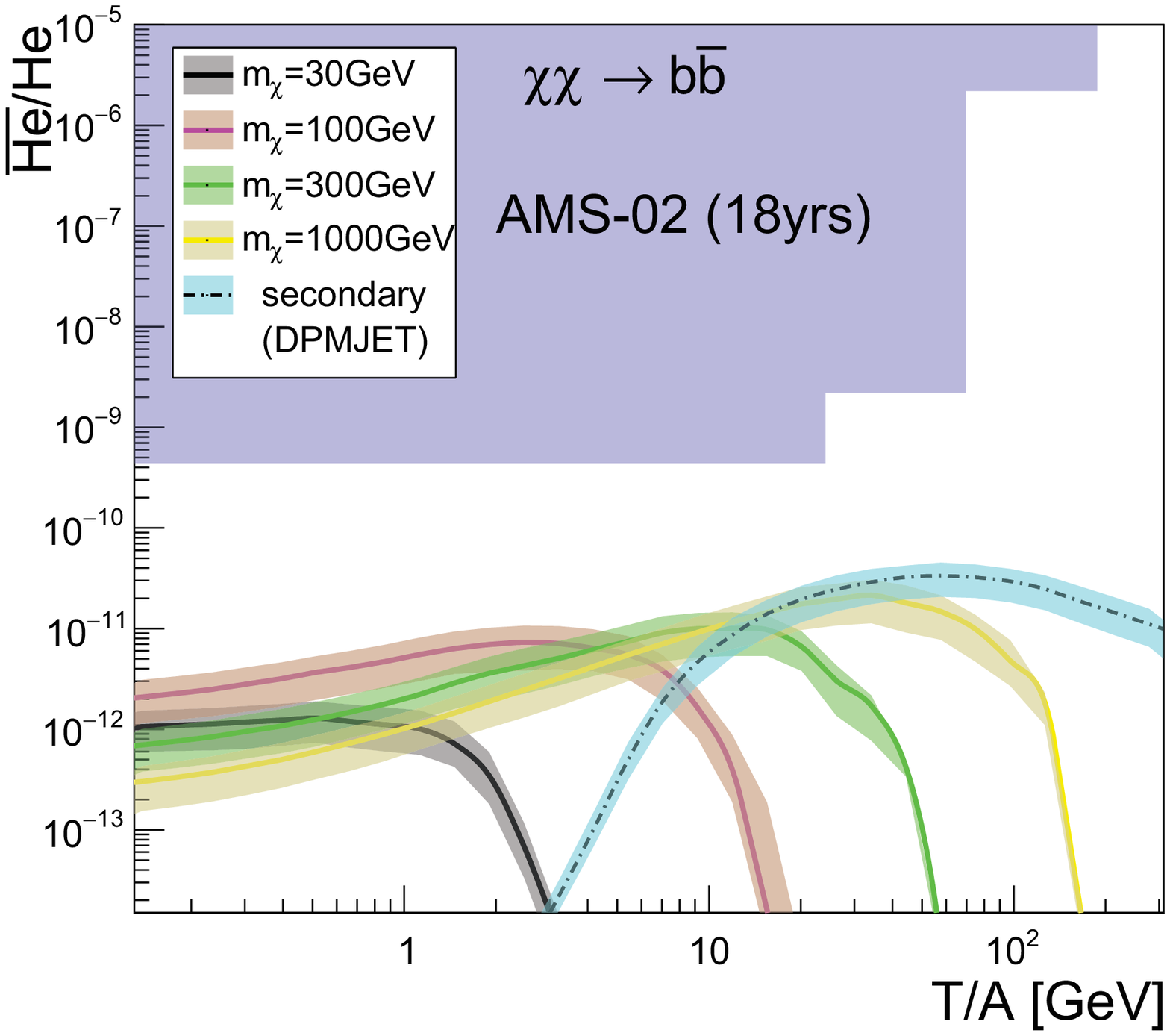}
	\includegraphics[width=0.32\textwidth]{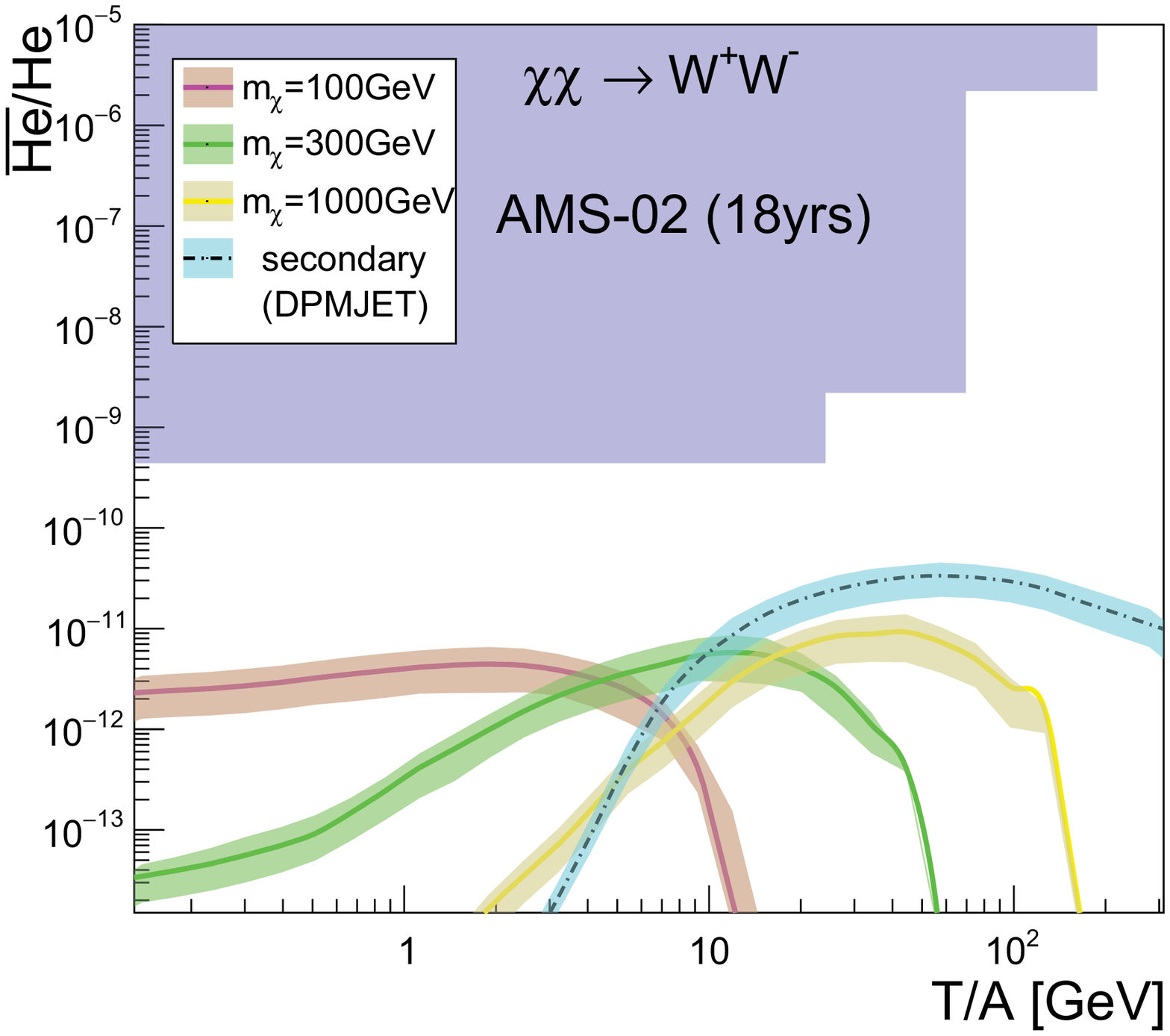}
	\caption{
		Upper panels) 
		maximal flux ratios  $\overline{\text{He}}/\text{He}$ from DM annihilation 
		under the constraints from the AMS-02 $\bar p$ data, together with  the secondary $\Hebar$
		backgrounds. The $pp$-collision cross section and the coalescence momentum $\pHebar$ are 
		determined using the MC event generator \EPOSLHC. The error bands indicate the uncertainty in the
		coalescence momentum. The results are obtained by adopting the ``MED''
		propagation model and the ``Isothermal'' DM profile.
		The blue shaded regions represent the detection sensitivity
		of AMS-02  at  $95\%$~C.L., after 18 years of data collection.
		Lower panels)
		the same as the upper panels, but based on the  event generator \DPMJET.
	}
	\label{fig:antihelium-flux-ratio}
\end{figure}

A more realistic estimation of the prospective 18-year
$^3\overline{\textrm{He}}$ detecting sensitivity of AMS-02 after
considering the contamination of $\textrm{He}$ was given in terms of
$\overline{\textrm{He}}/\mathrm{He}$ flux ratio in
Ref.~\cite{Kounine:2010js}, where He stands for $^3\mathrm{He}$+
$^4\mathrm{He}$.
In Fig.~\ref{fig:antihelium-flux-ratio}, we  show the predicted
$\overline{\textrm{He}}/\mathrm{He}$ in the ``MED'' propagation model
with the ``Isothermal'' DM profile for various DM annihilation
channels, DM particle masses, and coalescence momenta using
event generators {\tt EPOS-LHC} and {\tt DPMJET}.
We find that the events which can be observed are likely to have kinetic energy
$T/A \gtrsim 10$~GeV, i.e.,  $T\gtrsim 30$~GeV for $\Hebar$, and are dominantly arising
from the secondary backgrounds.
Recently, the AMS-02 collaboration has reported preliminary hints of
antihelium events~\cite{Ting:2016}.  For instance, a candidate event
of $^3\overline{\textrm{He}}$ with momentum $40.3 \pm 2.9$ GeV is
shown.  Such a event corresponds to a kinetic energy per nucleon
$T/A \approx12.5\pm 1.0$~GeV. As can be seen from the upper panels
of Fig.~\ref{fig:antihelium-flux-ratio}, it is close to the overlap region between
the AMS-02 sensitivity and the prediction from the secondary production.
Thus the energy of the candidate event is consistent with the secondary 
$\Hebar$ prediction in the most optimistic case. From Fig.~\ref{fig:antihelium-flux-ratio},
it is also evident that DM-interaction induced $\Hebar$ are unlikely to be
observed by AMS-02.

\begin{figure}[ht] \includegraphics[width=0.64\textwidth]{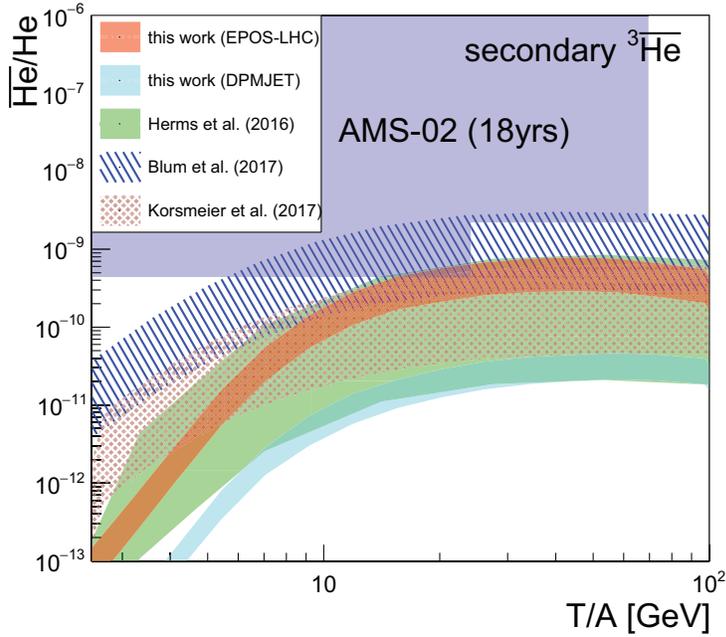} 
		\caption{
				A comparison of the secondary contribution to  the flux ratio 
				$\Hebar/\text{He}$ obtained in this work 
				with that from the previous work by 
				Herms, et.~al.~\cite{Herms:2016vop},
				Blum, et.~al.~\cite{Blum:2017qnn} and
				Korsmeier, et.~al.~\cite{Korsmeier:2017xzj}.
				See text for detailed discussions.
		}
 \label{fig:comparison}
 \end{figure}

For a comparison with the previous work in the literature,
in \fig{fig:comparison}, our results are compared with a selection of
previous work related to the projection of the  CR $\Hebar$ flux.
The analysis in Ref.~\cite{Herms:2016vop} considered the constraints
form the AMS-02 antiproton data, and used the coalescence model with
the \DPMJET~ event generator to simulate $\Hebar$ formation. However,
the coalescence momentum $\pHebar$ was inferred from the vale of
$\pDbar$, which is quite large $\pHebar \approx 311$~MeV and leads to
the conclusion that CR $\Hebar$ is within the reach of AMS-02. On the
country, in our work, we use the value $\pHebar =212^{+10}_{-13}$~MeV 
directly from fitting the ALICE data.  Consequently, in our work the predicted 
CR $\Hebar$ flux using \DPMJET~ is an order of magnitude lower.
In Ref.~\cite{Blum:2017qnn}, the coalescence parameter $B_3$ was
estimated using the ALICE data and the Hanbury-Brown-Twiss (HBT)
   two-particle-correlation measurements. The obtained value of $B_3$
is in a wide range $(2-20)\times 10^{-4}~\text{GeV}^4$. Consequently,
the predicted $\Hebar$ flux can be much larger than that from other
approaches. However, as shown by the ALICE data
(see \fig{fig:b3_bestfit}), the value of $B_3$ shows a significant
$p_T$ dependence which was not reproduced in the HBT approach
in~\cite{Blum:2017qnn}. Only in the highest $p_T$ bin the value of
$B_3$ can reach $20 \times 10^{-4}~\text{GeV}^4$. For lower $p_T$
bins, the corresponding $B_3$ can be smaller by an order of magnitude.
Note that the $p_T$ dependence of $B_3$ is correctly reproduced by the
all MC event generators considered in this work.
In Ref.~\cite{Korsmeier:2017xzj}, the maximal $\Hebar$ flux from DM
annihilation was discussed with the AMS-02 antiproton constraints
taken directly from Ref.~\cite{Cuoco:2016eej}. The analysis of \cite{Korsmeier:2017xzj} 
used the coalescence model and the
analytic relation between $\pHebar$ and $B_3$ in the isotropic limit,
which again cannot reproduced the $p_T$ dependence of $B_3$. To be
conservative, the value of $\pHebar$ was set in the range
$160-248$~MeV, which leads to large uncertainties in the predicted
maximal $\Hebar$ flux.
The $\Hebar$ flux from DM was also discussed recently in light of the
preliminary antihelium measurements by
AMS-02~\cite{Coogan:2017pwt}. However, neither the antiproton
constriants nor the background contributions was considered in
their analysis.

\section{Conclusions}\label{sec:conclusion}

In summary, motivated partly by the recent progresses made by AMS-02
in searching for heavier anti-nuclei, we have discussed the prospect
of detecting $^{3}\overline{\text{He}}$ in the AMS-02 experiment under
the constraints from the AMS-02 antiproton data.
We have updated the upper limits on DM annihilation cross sections
from the AMS-02 $\bar{p}/p$ ratio, and then used the results to set
limits on the $^3\overline{\textrm{He}}$ flux and number of events
which could be observed by AMS-02 in the whole lifetime of data
taking.
We have used the coalescence model to simulate the production of
$^3\overline{\textrm{He}}$ from DM annihilation on an event-by-event
basis, and used the \texttt{GALPROP} code to calculate the propagation
of $^3\overline{\textrm{He}}$ in the interstellar medium.  The results
show that with very optimistic estimates of detection efficiency and
acceptance, and a relatively large coalescence momentum, CR antihelium
is within the sensitivity of the AMS-02 experiment with a whole
lifetime of data taking.
The number of events can reach $\mathcal{O}(1)$, depending on the
value of the coalescence momentum.
We have also shown that the events which can be detected by AMS-02 are
likely to have kinetic energy $T \gtrsim 30$~GeV and dominantly arise
from secondary backgrounds rather than DM annihilation.

\section*{Acknowledgments}
We are grateful to Alejandro Ibarra for useful discussions on 
the prospect of detecting antihelium at AMS-02.
This work is supported in part by  the NSFC under Grants
No.~U1738209,
No.~11851303,
No.~11825506,
No.~11821505,
the National Key R\&D Program of China 
No.~2017YFA0402204, and
the CAS key research program
No.~XDB23030100,
No.~QYZDY-SSW-SYS007.
\bibliographystyle{arxivref}
\bibliography{Hebar3ref}

\end{document}